\documentclass[twocolumn,useAMS,usenatbib]{mn2e} \usepackage{natbib}
\usepackage{amsmath,latexsym,amssymb} \usepackage{epsfig,graphicx} \topmargin-1cm
\usepackage{float}
\usepackage{fixltx2e}
\graphicspath{ {./psubfigsv1/}}
\def\reff@jnl#1{{\rm#1\/}}

\def\aj{\reff@jnl{AJ}}                  
\def\araa{\reff@jnl{ARA\&A}}            
\def\apj{\reff@jnl{ApJ}}                        
\def\apjl{\reff@jnl{ApJ}}               
\def\apjs{\reff@jnl{ApJS}}              
\def\apss{\reff@jnl{Ap\&SS}}            
\def\aap{\reff@jnl{A\&A}}               
\def\aapr{\reff@jnl{A\&A~Rev.}}         
\def\aaps{\reff@jnl{A\&AS}}             
\def\baas{\reff@jnl{BAAS}}              
\def\jcap{\reff@jnl{JCAP}}              
\def\jrasc{\reff@jnl{JRASC}}            
\def\memras{\reff@jnl{MmRAS}}           
\def\mnras{\reff@jnl{MNRAS}}            
\def\physrep{\reff@jnl{Phys.Rep.}}
\def\pra{\reff@jnl{Phys.Rev.A}}         
\def\prb{\reff@jnl{Phys.Rev.B}}         
\def\prc{\reff@jnl{Phys.Rev.C}}         
\def\prd{\reff@jnl{Phys.Rev.D}}         
\def\prl{\reff@jnl{Phys.Rev.Lett}}      
\def\pasp{\reff@jnl{PASP}}              
\def\pasj{\reff@jnl{PASJ}}              
\def\skytel{\reff@jnl{S\&T}}            
\def\solphys{\reff@jnl{Solar~Phys.}}    
\def\sovast{\reff@jnl{Soviet~Ast.}}     
\def\ssr{\reff@jnl{Space~Sci.Rev.}}     
\def\nat{\reff@jnl{Nature}}             

\newcommand{\hmpc}{\ensuremath{h^{-1}\mathrm{Mpc}}}
\newcommand{\hkpc}{\ensuremath{h^{-1}\mathrm{kpc}}}
\newcommand{\hMsun}{\ensuremath{h^{-1}M_{\odot}}}
\newcommand{\msubhalo}{\ensuremath{M_\text{subhalo}}}

\newcommand{\beq}{\begin{equation}}
\newcommand{\eeq}{\end{equation}}
\newcommand{\beqa}{\begin{eqnarray}}
\newcommand{\eeqa}{\end{eqnarray}}

\title[Shapes Alignments]{Galaxy Shapes and Intrinsic Alignments in the MassiveBlack-II Simulation}
\author[Tenneti et al.]
{Ananth Tenneti$^1$\thanks{\tt vat@andrew.cmu.edu},
Rachel Mandelbaum$^1$\thanks{\tt rmandelb@andrew.cmu.edu},
Tiziana Di Matteo$^1$\thanks{\tt tiziana@phys.cmu.edu},
Yu Feng$^1$, 
\newauthor Nishikanta Khandai$^2$
\\$^1$McWilliams Center for Cosmology, Department of Physics, Carnegie Mellon University, Pittsburgh, PA 15213, USA
\\$^2$Department of Physics, Brookhaven National Laboratory, Upton, NY 11973, USA }

\date{\today}
\begin{document}
\maketitle

\begin{abstract}
  The intrinsic alignment of galaxy shapes with the large-scale
  density field is a contaminant to weak lensing measurements, as well
  as being an interesting signature of galaxy formation and evolution
  (albeit one that is difficult to predict theoretically).  Here we
  investigate the shapes and relative orientations of the stars and
  dark matter of halos and subhalos (central and satellite) extracted
  from the MassiveBlack-II simulation, a state-of-the-art high
  resolution hydrodynamical cosmological simulation which includes
  stellar and AGN feedback in a volume of $(100$\hmpc$)^3$. We
  consider redshift evolution from $z=1$ to $0.06$ and mass evolution
  within the range of subhalo masses, $10^{10} -6.0 \times
  10^{14.0}$\hMsun.  The shapes of the dark matter distributions are
  generally more round than the shapes defined by stellar matter. The
  projected root-mean-square (RMS) ellipticity per component for
  stellar matter is measured to be $e_\text{rms} = 0.28$ at $z=0.3$
  for $\msubhalo> 10^{12.0}$\hMsun, which compares favourably with observational measurements. We find that the shapes
  of stellar and dark matter are more round for less massive subhalos
  and at lower redshifts. By directly measuring the relative
  orientation of the stellar matter and dark matter of subgroups, we
  find that, on average, the misalignment between the two components is larger for
  less massive subhalos. The mean misalignment angle varies from $\sim
  30^{\circ}-10^{\circ}$ for $M \sim 10^{10} - 10^{14}
  \hMsun$ and shows a weak dependence on redshift. We also
  compare the misalignment angles in central and satellite subhalos at
  fixed subhalo mass, and find that centrals are more misaligned than
  satellites. We present fitting formulae for the shapes of dark and stellar matter in subhalos and also the probability distributions of
  misalignment angles.
\end{abstract}


\section{Introduction} \label{S:intro}

Weak gravitational lensing is a useful probe to constrain cosmological
parameters since it is sensitive to both luminous and dark matter
\citep{{2002PhRvD..65b3003H},{2004PhRvD..70l3515B},{2004PhRvD..69h3514I},{2004ApJ...601L...1T},{2004ApJ...600...17B},{2010GReGr..42.2177H}}. In
particular, weak lensing surveys can be used to probe theories of
modified gravity and provide constraints on the properties of dark
matter and dark energy
\citep{{2006astro.ph..9591A},{2013PhR...530...87W}}. Many
upcoming surveys like Large Synoptic Survey Telescope(LSST)\footnote{http://www.lsst.org/lsst/} and Euclid\footnote{http://sci.esa.int/euclid/} aim to determine the constant
and dynamical parameters of the dark energy equation of state to a
very high precision using weak lensing.

However, constraining cosmological parameters with sub-percent errors in future
cosmological survey requires the systematic errors to be well below
those in typical weak lensing measurements with current datasets. The intrinsic shapes and orientations of galaxies are not random but correlated with each other and the underlying density field. This is known as intrinsic galaxy alignments. The
intrinsic alignment (IA) of galaxy shapes with the underlying density
field is an important theoretical uncertainty that contaminates weak
lensing measurements
\citep{{2000MNRAS.319..649H},{2000ApJ...545..561C},{2002MNRAS.335L..89J},{2004PhRvD..70f3526H}}. Accurate
theoretical predictions of IA through analytical models and $N$-body
simulations \citep{2004PhRvD..70f3526H,2006MNRAS.371..750H,2010MNRAS.402.2127S,2013MNRAS.436..819J} in the $\Lambda$CDM paradigm is
complicated by the absence of baryonic physics, which we expect to be
important given that the alignment of interest is that of the
observed, baryonic component of galaxies. So, we either need simulations that include the physics of galaxy formation or $N$-body simulations with rules for galaxy shapes and alignments.

Proposed analysis methods to remove IA from weak lensing measurements
either involve removing considerable amount of cosmological information (which
requires very accurate redshift information; nulling methods: 
\citealt{2008A&A...488..829J, 2009A&A...507..105J}), or involve
marginalizing over parametrized models of how the intrinsic alignments
affect observations as a function of scale, redshift and galaxy type
\citep[e.g., ][]{2007NJPh....9..444B,2010A&A...523A...1J,2012JCAP...05..041B}. The
simultaneous fitting method, with a relatively simple intrinsic
alignments model, was used for a tomographic cosmic shear analysis of
CFHTLenS data \citep{2013MNRAS.432.2433H}. The latter methods, while
preserving more cosmological information than nulling methods, can
only work correctly if there is a well-motivated intrinsic alignments
model as a function of galaxy properties.  Existing candidates for the
intrinsic alignment model to be used in such an approach include the
linear alignment model \citep{2004PhRvD..70f3526H} or simple
modifications of it (e.g., using the nonlinear power spectrum:
\citealt{2007NJPh....9..444B}), $N$-body simulations populated with galaxies and
stochastically misaligned with halos in a way that depends on galaxy
type \citep{2006MNRAS.371..750H}, and the halo model
\citep{2010MNRAS.402.2127S}, which includes rules for how central and
satellite galaxies are intrinsically aligned. 

In this study, we use the large volume, high-resolution hydrodynamic
simulation, MassiveBlack-II \citep{2014arXiv1402.0888K}, which
includes a range of baryonic processes to directly study the shapes
and alignments of galaxies. In particular, we measure directly the
shapes of the dark and stellar matter components of halos and subhalos
(modeled as ellipsoids in three dimensional space). We examine how
shapes evolve with time and as a function of halo/subhalo
mass. Previous work used $N$-body simulations and analytical modeling
to study triaxial shape distributions of dark matter halos as a
function of mass and their evolution with redshift
\citep{{2005ApJ...618....1H},{2006MNRAS.367.1781A},{2005ApJ...632..706L},{2012JCAP...05..030S}}.
More recently, hydrodynamic cosmological simulations have also been
used to study the effects of baryonic physics on the shapes of dark
matter halos
\citep{{2005ApJ...627L..17B},{2006EAS....20...65K},{2010MNRAS.405.1119K},{2013MNRAS.429.3316B}}.
Here, using a high-resolution hydrodynamic simulation in a large
cosmological volume that incorporates the physics of star formation
and associated feedback as well as black hole accretion and AGN
feedback, we focus on measuring directly the shapes of the stellar
components of galaxies and examine the misalignments between stars and
dark matter in galaxies (central and satellite). We also measure the
projected (2D) shapes for comparison with observations. This study is
important because the measured intrinsic alignments of galaxies are
related to the projected shape correlations of the stellar component
of subgroups (galaxies) by the density-ellipticity and
ellipticity-ellipticity correlations \citep{2006MNRAS.371..750H}. By
measuring the projected ellipticities of the stellar and dark matter
component of simulated galaxies, we can attempt to understand the
differences between these two. In addition, we can do a basic
comparison of the stellar components with observational results, and
validate the realism of the simulated galaxy population.

Another aspect of the problem that we consider in this paper is the
relative orientation of the stellar component of the halo with its
dark matter component. Many dark matter-only simulations have
illustrated that dark matter halos exhibit large-scale intrinsic
alignments \citep[e.g.,
][]{2002A&A...395....1F,2005ApJ...618....1H,2006MNRAS.370.1422A,2006MNRAS.371..750H},
but the prediction of galaxy intrinsic alignments from halo intrinsic
alignments requires a statistical understanding of the relationship
between galaxy and halo shapes. To date, there has been no direct
measurement of galaxy versus halo misalignment with a large
statistical sample of galaxies through hydrodynamic
simulations. Recently, \cite{2014arXiv1402.1165D} studied the alignment between the spin of galaxies and their host filament direction using a hydrodynamical cosmological simulation of box size $100\hmpc$. Studies of misalignment based on SPH simulations of
smaller volumes detected misalignments between the baryonic and dark
matter component of halos
\citep{{2003MNRAS.346..177V},{2005ApJ...628...21S},{2010MNRAS.405..274H},{2011MNRAS.415.2607D}}. These
studies considered the correlation of spin and angular momentum of the
baryonic component with dark matter. The spin correlations are
arguably more relevant for the intrinsic alignments of spiral galaxies
\citep{2004PhRvD..70f3526H}, whereas the observed intrinsic alignments
in real galaxy samples are dominated by red, pressure-supported,
elliptical galaxies \citep{2011MNRAS.410..844M,2011A&A...527A..26J};
hence a study of the correlation of projected shapes is more relevant
for the issue of weak lensing contamination. However, to make precise
predictions based on the halo or subhalo mass at different redshifts,
we need a hydrodynamic simulation of very large volume and high
resolution. The MassiveBlack-II SPH simulation meets those
requirements, making it a good choice for this kind of study.

Others arrived at constraints on misalignments using $N$-body
simulations and calibrating the misalignments by adopting a simple
parametric form to agree with observationally detected shape
correlation functions
\citep{{2009RAA.....9...41F},{2009ApJ...694..214O}}. There are also
studies of the alignment of a central galaxy with its host halo where
it is assumed that the satellites trace the dark matter distribution
\citep[e.g., ][]{2008MNRAS.385.1511W}. By using hydrodynamic
simulations, we can directly calculate the misalignment distributions
for all galaxies as a function of halo mass and cosmic
time. Resolution of the galaxies into centrals and satellites also
helps to understand the effect of local environment.

This paper is organized as follows. In Section~\ref{S:methods}, we
describe the SPH simulations used for this work and the methods used
to obtain the shapes and orientations of groups and subgroups. In
Section~\ref{S:shapes}, we give the axis ratio distributions of dark
matter and stellar matter of subgroups. In
Section~\ref{S:misalignments}, we show our results for misalignments
of the stellar component of subgroups with their host dark matter
subgroups. In Section~\ref{S:censat} we compare the shape
distributions and misalignment angle between centrals and
satellites. Finally, we summarize our conclusions in
Section~\ref{S:conclusions}. The functional forms for our results are
provided in the Appendix.

\section{Methods}\label{S:methods}

\subsection{MassiveBlack-II Simulation}
\begin{figure*}
\begin{center}
$\begin{array}{c@{\hspace{0.1in}}c}
\includegraphics[width=\textwidth,angle=0]{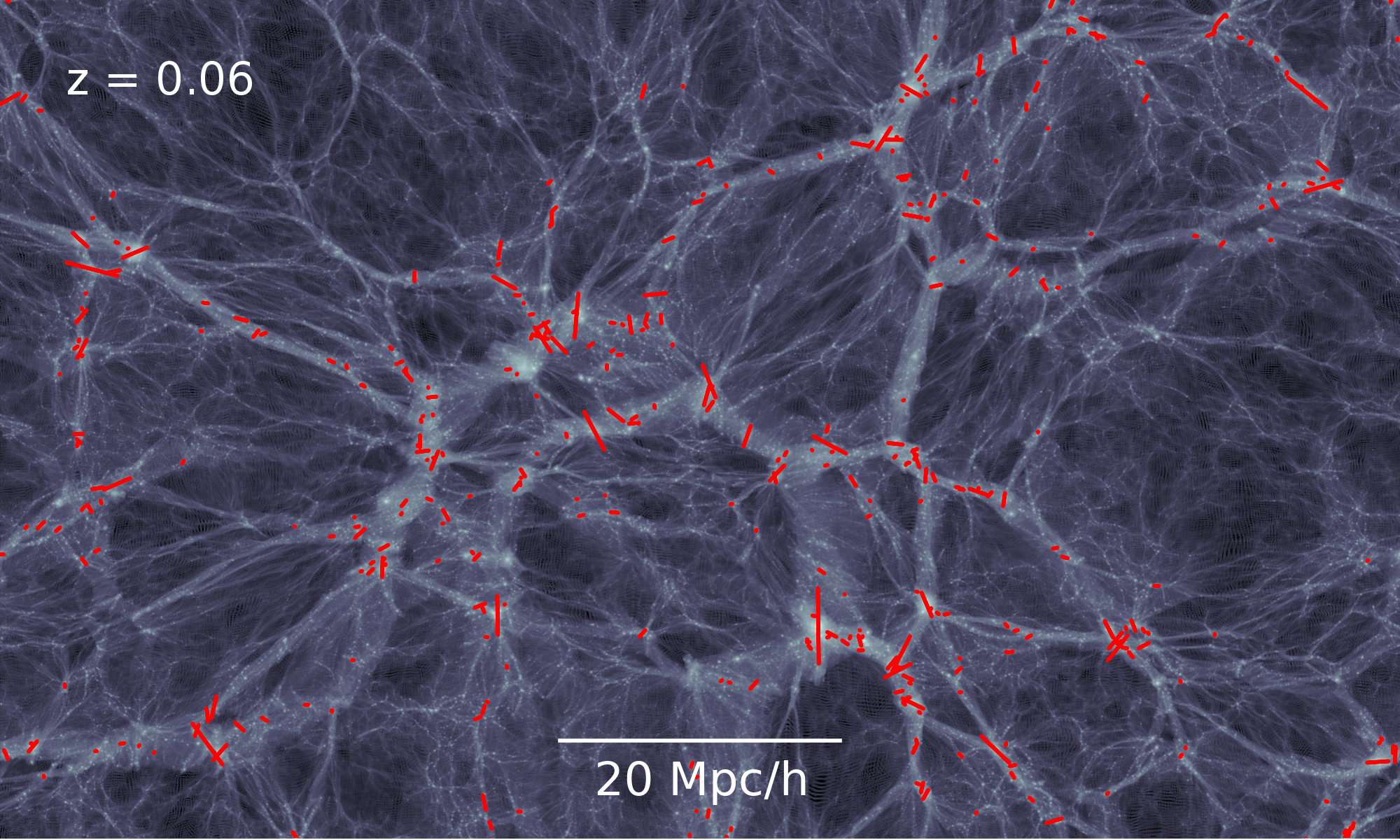} \\
\includegraphics[width=2.3in,height=2.3in,angle=0]{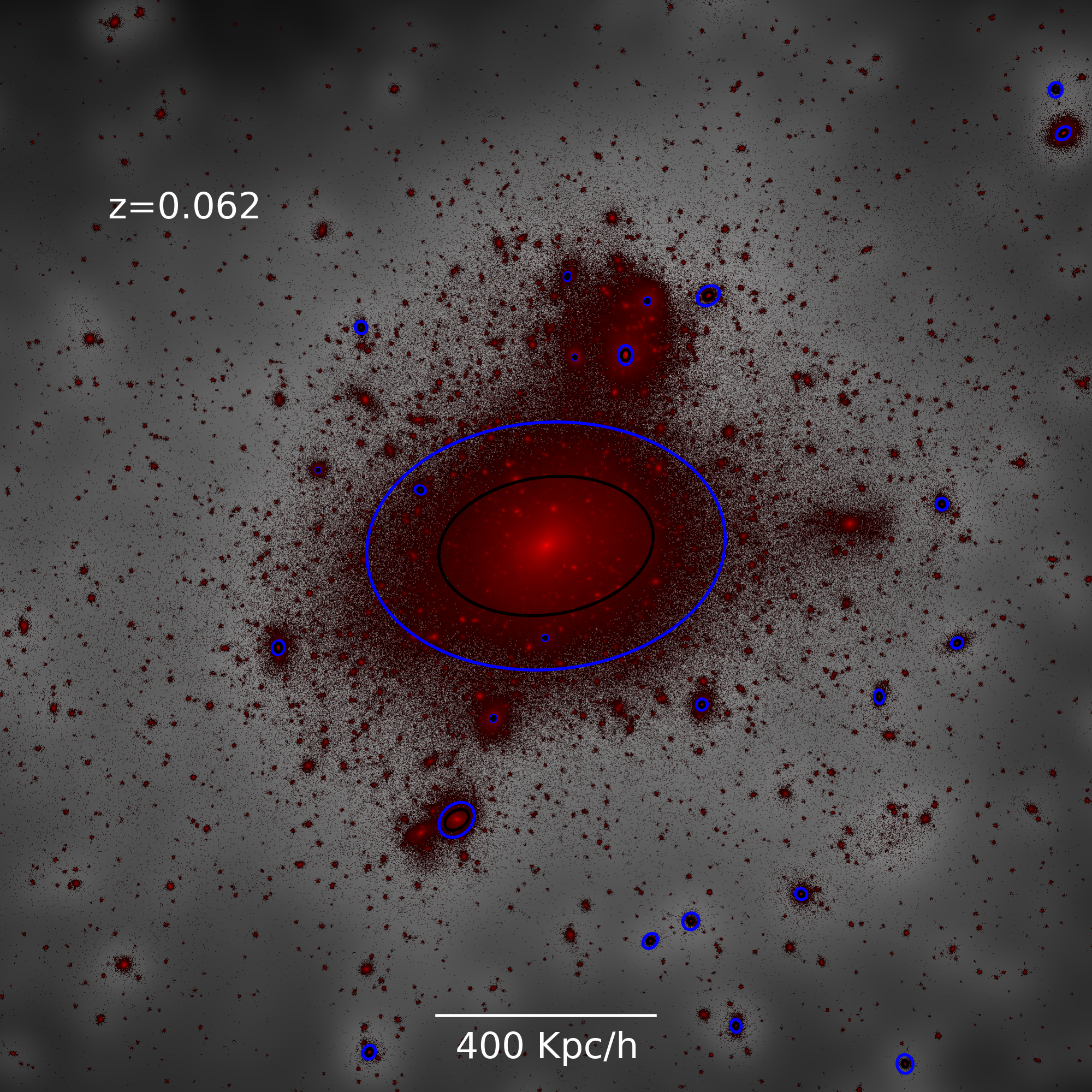}\
\includegraphics[width=2.3in,height=2.3in,angle=0]{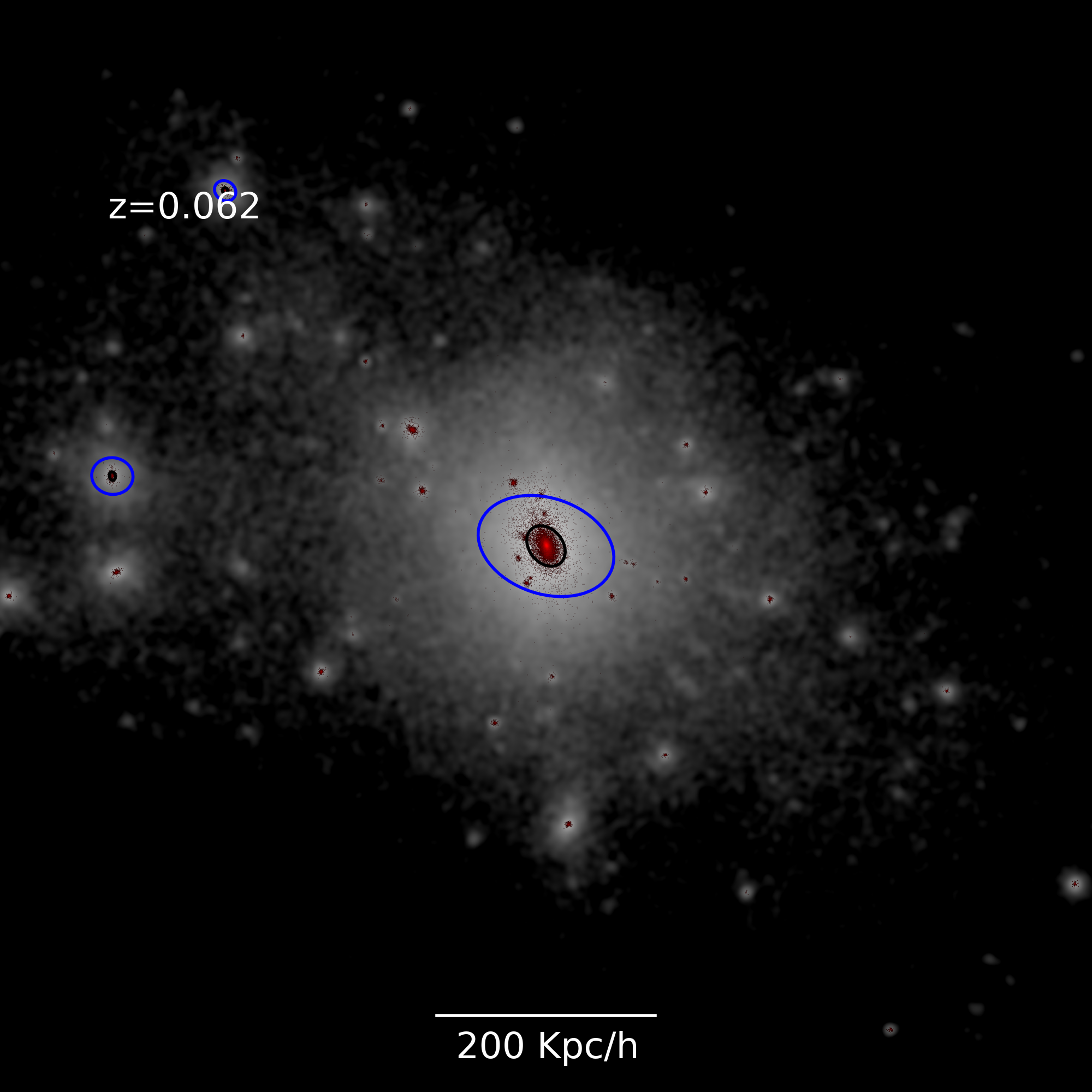}\ 
\includegraphics[width=2.3in,height=2.3in,angle=0]{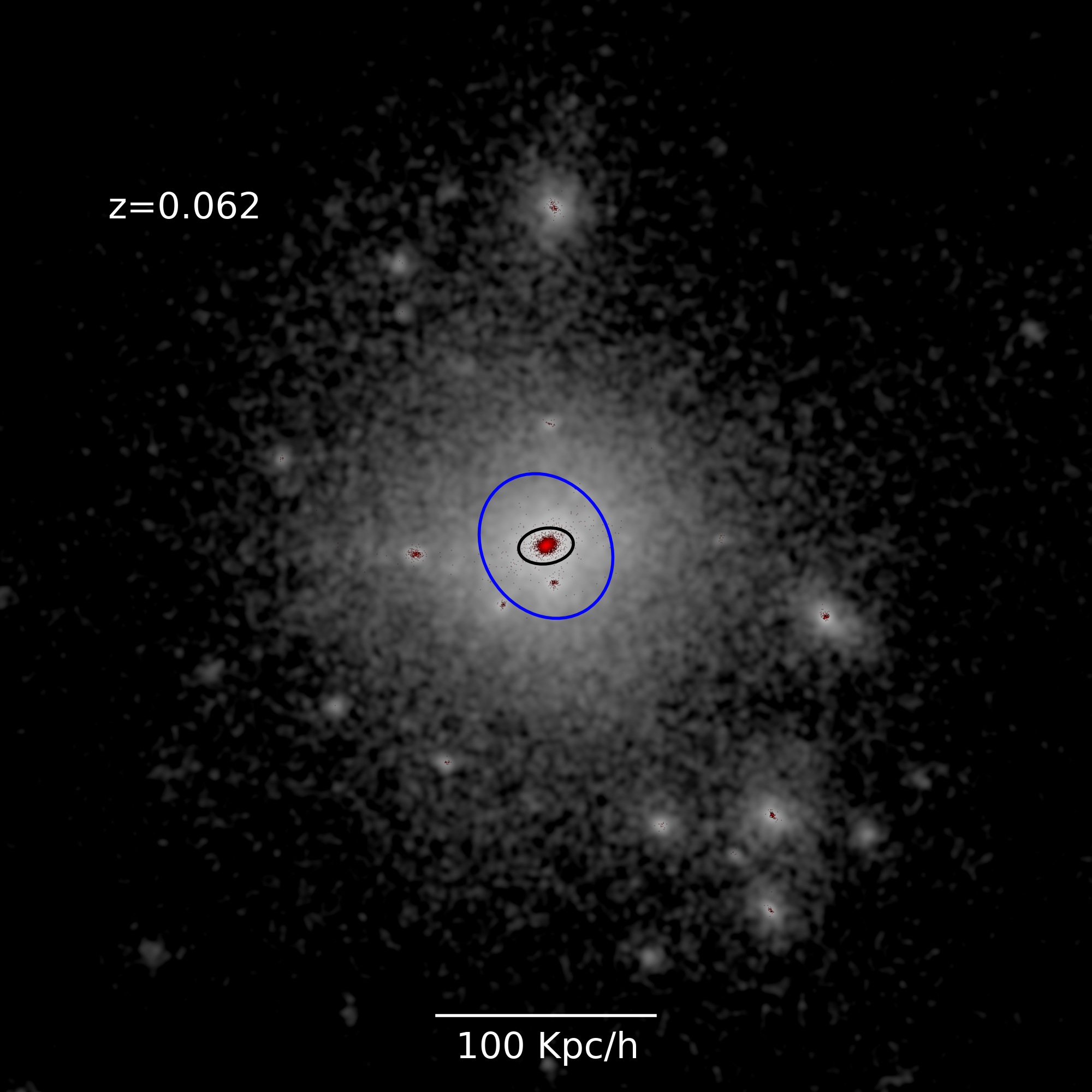}\\
\end{array}$
\caption{\label{F:fig112}{\em Top:} Snapshot of the MBII simulation in a slice of
  thickness $2$\hmpc\ at redshift $z = 0.06$. The bluish-white colored region represents the density of the dark matter distribution and the red lines show the direction of the major axis of ellipse for the projected shape defined by the stellar component. {\em Bottom Left:} Dark matter (shown in gray) and stellar matter (shown in red)
  distribution in the most massive group at $z = 0.06$ of mass $7.2 \times
  10^{14}$\hMsun. The blue and red ellipses show the projected shapes of dark matter and stellar matter of subhalos respectively. {\em Bottom Middle:} Dark matter and stellar matter distribution in a
  group of mass $3.8 \times 10^{12}$\hMsun. {\em Bottom Right:} Dark matter and
  stellar matter distribution in a group of mass $1.1 \times 10^{12}$\hMsun.}
\end{center}
\end{figure*}

\begin{figure}
\begin{center}
\includegraphics[height=\columnwidth,angle=0]{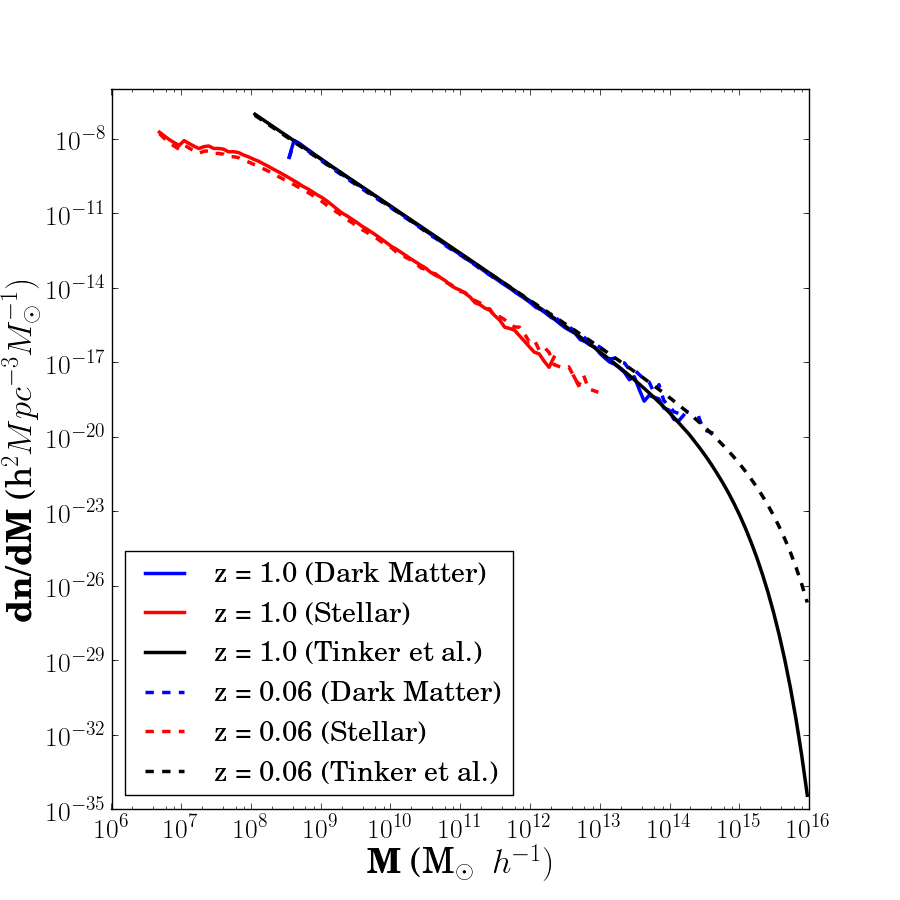}
\caption{\label{F:fig_mf} Dark matter and stellar mass function for
  FOF groups (halos) at $z = 0.06,
  1.0$, compared with the SO-based prediction from \protect\cite{2008ApJ...688..709T} generated with $\Delta = 0.75$.}
\end{center}
\end{figure}

We use the MassiveBlack-II (MBII) simulation to measure shapes and
alignments of dark matter and stellar components of halos and
subhalos. MBII is a state-of-the-art high resolution, large volume,
cosmological hydrodynamic simulation of structure formation. An
extensive description of the simulation and major predictions for the
halo and subhalo mass functions, their clustering, the galaxy stellar mass
functions, galaxy spectral energy distribution and properties of the
AGN population is presented by \cite{2014arXiv1402.0888K}. We refer the reader
to this publication for details on MBII and briefly summarize the major
relevant aspects here.

The MBII simulation was performed with the cosmological TreePM-Smooth
Particle Hydrodynamics (SPH) code {\sc p-gadget}. It is a hybrid
version of the parallel code, {\sc gadget2}
\citep{2005MNRAS.361..776S} that has been upgraded to run on Petaflop
scale supercomputers.  In addition to gravity and SPH, the {\sc
  p-gadget} code also includes the physics of multi-phase ISM model
with star formation \citep{2003MNRAS.339..289S}, black hole accretion
and feedback \citep{2005MNRAS.361..776S,2012ApJ...745L..29D}.
Radiative cooling and heating processes are included \citep[as
in][]{1996ApJS..105...19K}, as is photoheating due to an imposed
ionizing UV background.  The interstellar medium (ISM), star formation
and supernovae feedback as well as black hole accretion and associated
feedback are treated by means of previously developed sub-resolution
models. In particular, the multiphase model for star forming gas we
use, developed by \cite{Springel2003a}, has two principal ingredients:
(1) a star formation prescription and (2) an effective equation of
state (EOS).  A thermal instability is assumed to operate above a
critical density threshold $\rho_{\rm th}$, producing a two phase
medium consisting of cold clouds embedded in a tenuous gas at pressure
equilibrium. Stars form from the cold clouds, and short-lived stars
supply an energy of $10^{51}\,{\rm ergs}$ to the surrounding gas as
supernovae. This energy heats the diffuse phase of the ISM and
evaporates cold clouds, thereby establishing a self-regulation cycle
for star formation.  $\rho_{\rm th}$ is determined self-consistently
in the model by requiring that the EOS is continuous at the onset of
star formation. Stellar feedback in the form of stellar winds is also
included.  The prescription for black hole accretion and associated
feedback from massive black holes follows the one developed by
~\citet{DiMatteo2005, Springel2005a}. We represent black holes by
collisionless particles that grow in mass by accreting gas (at the
local dynamical timescale) from their environments.  If the accretion
rates reach the critical Eddington limit they are then capped at that
value.  A fraction $f$ (fixed to $5\%$ to fit the local black-hole
galaxy relations) of the radiative energy released by the accreted
material is assumed to couple thermally to nearby gas and influence
its thermodynamic state. Black holes merge when they approach the
spatial resolution limit of the simulation \citep{Springel2003a}

MBII contains $N_\mathrm{part} = 2\times 1792^{3}$ dark matter and gas
particles in a cubic periodic box of length $100$\hmpc\ on a side,
with a gravitational smoothing length $\epsilon = 1.85$\hkpc\ in
comoving units. A single dark matter particle has a mass $m_{DM} = 1.1\times 10^{7}\hMsun$ and the initial mass of a gas particle is $m_{gas} = 2.2\times 10^{6}\hMsun$. The cosmological parameters used in the simulation are
as follows: amplitude of matter fluctuations $\sigma_{8} = 0.816$,
spectral index $\eta_{s} = 0.96$, mass density parameter $\Omega_{m} =
0.275$, cosmological constant density parameter $\Omega_{\Lambda} =
0.725$, baryon density parameter $\Omega_{b} = 0.046$, and Hubble
parameter $h = 0.702$ as per WMAP7 \citep{2011ApJS..192...18K}.

Fig.~\ref{F:fig112} shows snapshots of the MBII simulation with
dark matter and stellar matter distributions at redshift $z =
0.06$. From the top figure, we can see the formation of cosmic web
with galaxies extending over the whole length of the simulation
volume. The bluish-white colored region in the figure represents the density of the dark matter distribution and the red lines show the direction of the major axis of ellipse for the projected shape defined by the stellar component. The figures in the bottom panel, which are zoomed snapshots of
individual halos of different masses, show the density distribution of
dark matter and stellar matter. The over plotted blue and red ellipses depict the projected shapes of dark matter and stellar matter of subhalos respectively.

To generate group catalogs of particles in the simulation, we used the
friends of friends (FoF) group finder algorithm \citep{1985ApJ...292..371D}. This algorithm
identifies groups on the fly using linking length of $0.2$ times the
mean interparticle separation. The mass of a halo is equal to the sum
of masses of all particles in the group. Fig.~\ref{F:fig_mf} shows the
dark matter and stellar mass functions for groups at redshifts $z =
1.0$ and $z = 0.06$. We find good agreement with the theoretical
prediction given in \cite{2008ApJ...688..709T} based on Spherical Overdensity (SO) approach. This gives an idea of
the mass range we are exploring by the use of this simulation. To
generate subgroup catalogs, the {\sc subfind} code
\citep{2001MNRAS.328..726S} is used on the group catalogs. The
subgroups are defined as locally overdense, self-bound particle
groups. Groups of particles are defined as subgroups when they have at
least $20$ gravitationally bound particles. A comparison between the
properties of halos and subhalos recovered using different halo and
subhalo finders can be found in \cite{2011MNRAS.415.2293K}, where it
is concluded that the properties of halos and subhalos, like mass,
position, velocity, two-point correlation returned by different
finders agree within error bars to each other. In all the discussions
in this paper, halos and subhalos are interchangeable for groups and
subgroups respectively.

\subsection{Determination of 3D and 2D shapes}\label{SS:shapedet}
Here we describe the method adopted to determine the shapes and orientations
of groups and subgroups for dark matter and stellar components. For each group and
subgroup, the dark matter and stellar shapes are determined by using the positions of dark
matter and star particles respectively. By using the positions of all particles of the
corresponding type, the halo and subhalo shapes in 3D are modelled as ellipsoids. For
projected shapes, the positions of particles of corresponding type projected onto the $XY$
plane are used to model the shapes as ellipses. We use the unweighted inertia tensor given by \\
\begin{equation} \label{eq:inertiatensor}
 I_{ij} = \frac{\sum_{n} m_{n}x_{ni}x_{nj}}{\sum_{n} m_{n}},
\end{equation}
where $m_{n}$ represents the mass of the $n^{th}$ particle and
$x_{ni},x_{nj}$ represent the position coordinates of the $n^{th}$
particle with $ 0 \leq i,j \leq 2$ for 3D and $0 \leq i,j \leq 1$ for
2D. It is to be noted that in this simulation, all particles of the
given type (either dark matter or star particle) have the same
mass. Hence the mass of a particle has no effect on the inertia
tensor. The inertia tensor can also be defined by weighting the
positions of particles by their luminosity instead of
mass. \cite{2012JCAP...05..030S} used the definition of reduced
inertia tensor and investigated the radial dependance of halo shapes
in the $N$-body simulation by considering only particles within a
given fraction of the virial radius. In this paper, we are only
concerned with the standard unweighted inertia tensor definition for
determining shapes and defer investigation of other definitions for a
future study.
 
Consider the 3D case. Let the eigenvectors of the inertia tensor be
${\hat{e}_{a},\hat{e}_{b},\hat{e}_{c}}$ and the corresponding eigenvalues be
${\lambda_{a},\lambda_{b},\lambda_{c}}$, where $\lambda_{a} > \lambda_{b} >
\lambda_{c}$. The eigenvectors represent the principal axes of the ellipsoids with the
lengths of the principal axes $(a,b,c)$ given by the square roots of
the eigenvalues 
$(\sqrt{\lambda_{a}},\sqrt{\lambda_{b}},\sqrt{\lambda_{c}})$. We now define
the 3D axis ratios as
\begin{equation} \label{eq:axisratios}
q = \frac{b}{a}, \,\, s = \frac{c}{a}
\end{equation}

In 2D, the eigenvectors are ${\hat{e}_{a}',\hat{e}_{b}'}$ with the corresponding
eigenvalues ${\lambda_{a}',\lambda_{b}'}$, where ${\lambda_{a}' > \lambda_{b}'}$. The lengths
of major and minor axes are $a' = \sqrt{\lambda_{a}'}$, $b' = \sqrt{\lambda_{b}'}$ with axis
ratio, $q' = b'/a'$ as defined before.

Our predictions from SPH simulations can be compared with those from
$N$-body simulations using the full 3D shapes, while the projected
shapes are useful for comparison with results from observational
data. In all our results, we used groups and subgroups with a minimum
of $1000$ dark matter and star particles each. We describe the
convergence tests performed to arrive at this cutoff in Section~\ref{SS:convshapes}.

\begin{figure*}
\begin{center}
$\begin{array}{c@{\hspace{0.5in}}c}
\includegraphics[width=3.2in,angle=0]{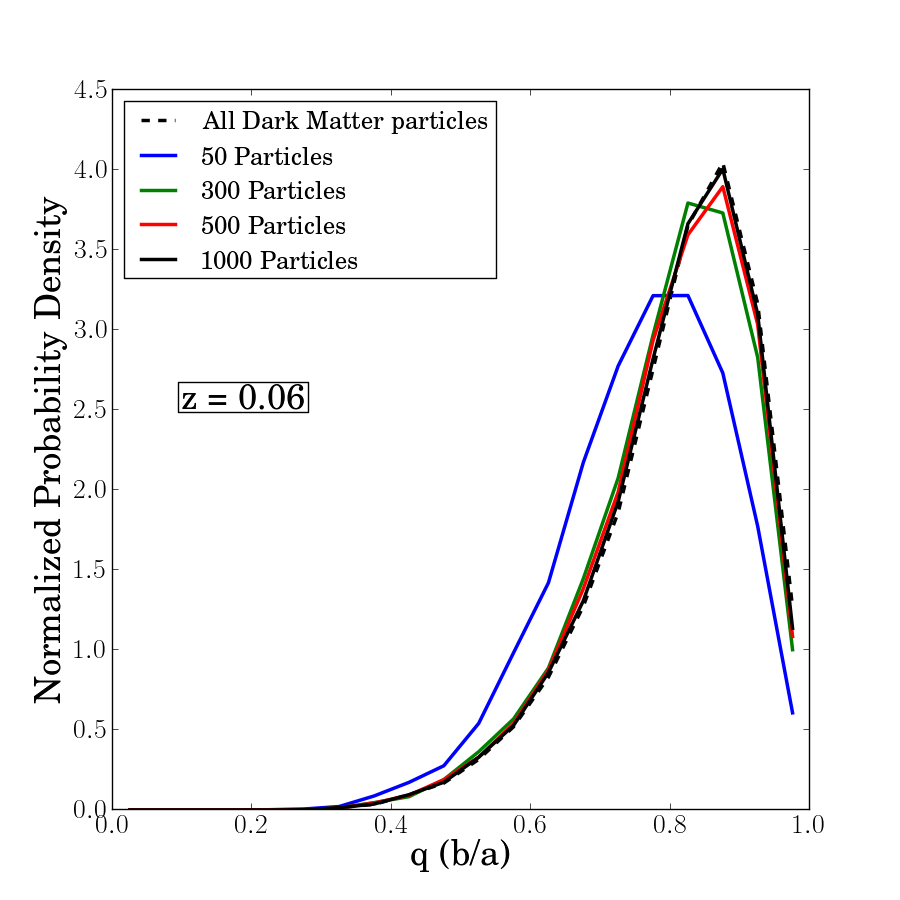} &
\includegraphics[width=3.2in,angle=0]{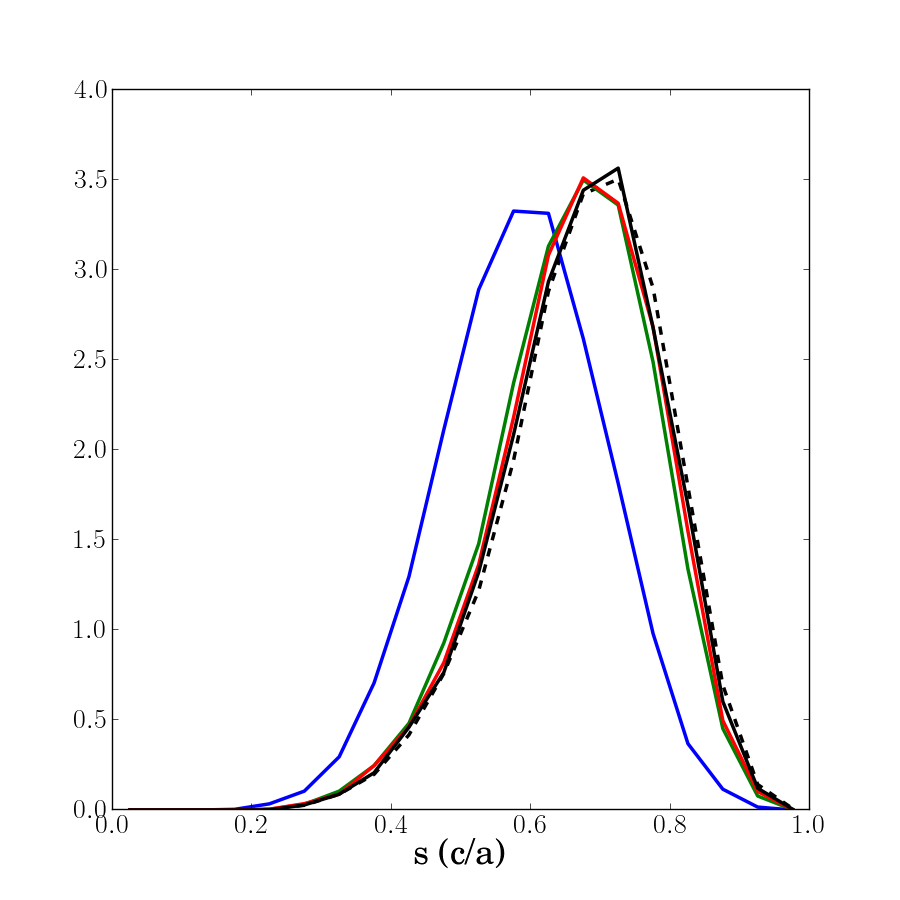}
\end{array}$
\caption{\label{F:fig_subsamp1000} Normalized histograms of axis ratios at $z=0.06$ showing
  a comparison between shapes determined by using all particles in the subhalo with
  those obtained using a random subsample of $50$, $300$, $500$ and
  $1000$ particles in the subhalo. {\em
    Left:} $q~(b/a)$; {\em Right:} $s~(c/a)$.}
\end{center}
\end{figure*}

\subsection{Convergence tests on axis ratios}\label{SS:convshapes}

\begin{figure}
\begin{center}
\includegraphics[height=\columnwidth,angle=0]{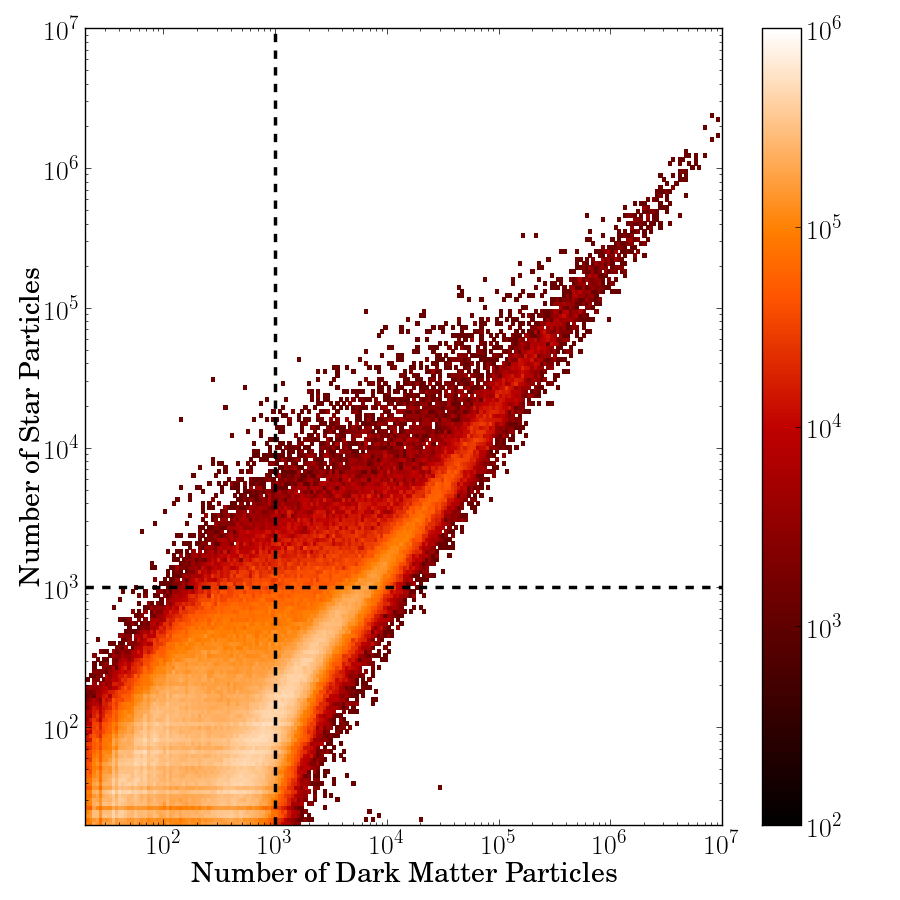}
\caption{\label{F:fig_contour}Distribution of the number of dark matter and star
  particles in subgroups at $z = 0.06$, where the colorbar indicates the number density of subhalos.}
\end{center}
\end{figure}

The reliability of statements about the shapes of matter distributions
depends on the number of particles used to trace those distributions.
Thus, we made a convergence test to fix the minumum number of particles
needed to measure shapes of halos and subhalos reliably. In
Fig.~\ref{F:fig_subsamp1000}, we show the histograms of shapes
measured using all the dark matter particles in a given subhalo, and
compared it with the histograms obtained by using a random subsample
of $50$, $300$, $500$ and $1000$ particles in the subhalo. This is
done in a mass range where we have enough subhalos with $>1000$
particles. The plots show that using a random subsample of $1000$
particles, we have a good convergence with the shapes determined using
all particles. The mean axis ratio, $\langle q \rangle$ is $0.83$ and
$\langle s \rangle$ is $0.70$ using all particles. $\langle q \rangle$
varies as $0.77,0.82,0.82,0.83,0.83$ using $50,300,500,1000$ particles
respectively. The corresponding values for $\langle s \rangle$ are
$0.60,0.68,0.69,0.70,0.70$. Although the mean axis ratios show good
convergence with $300$ or $500$ particles, from the plots we can see
that the histograms have not converged. Hence, we choose a minimum of
$1000$ particles for our analysis. In Figure~\ref{F:fig_contour}, we
show a contour plot of the number of dark matter particles and star
particles in subgroups at $z = 0.06$. The two different density peaks
in the contour plot are due to different dark matter to stellar mass
ratios in centrals and satellite subgroups. The right density peak
corresponds to central subhalos while the left one is for satellite
subgroups, which exhibit stripping of the dark matter subhalo and
hence fewer dark matter particles. The lines show a cutoff of $1000$
particles for dark matter and star particles. By choosing this cutoff,
we are excluding subhalos of low stellar to halo mass ratio in
subhalos around the low mass range $10^{10}-10^{11.5}$\hMsun. So in
this mass range, we are excluding a significant fraction of subhalos
with low stellar mass from our analysis. However, in the high mass
range, we are able to analyze a fair sample of subhalos.

\section{Shapes of dark matter and stellar matter of subgroups}\label{S:shapes}

In this section, we show the axis ratio distributions of the shapes of
dark matter and stellar matter component of halos and subhalos modeled
as ellipsoids as described in Section~\ref{SS:shapedet}. We
investigate their dependence on the mass range of subgroups and their
evolution with redshift. We also compare the relative axis ratio
distributions of dark matter and stellar matter in subhalos.
      
\subsection{3D axis ratio distributuions}

\begin{figure*}
\begin{center}
\includegraphics[width=1.0\textwidth,angle=0]{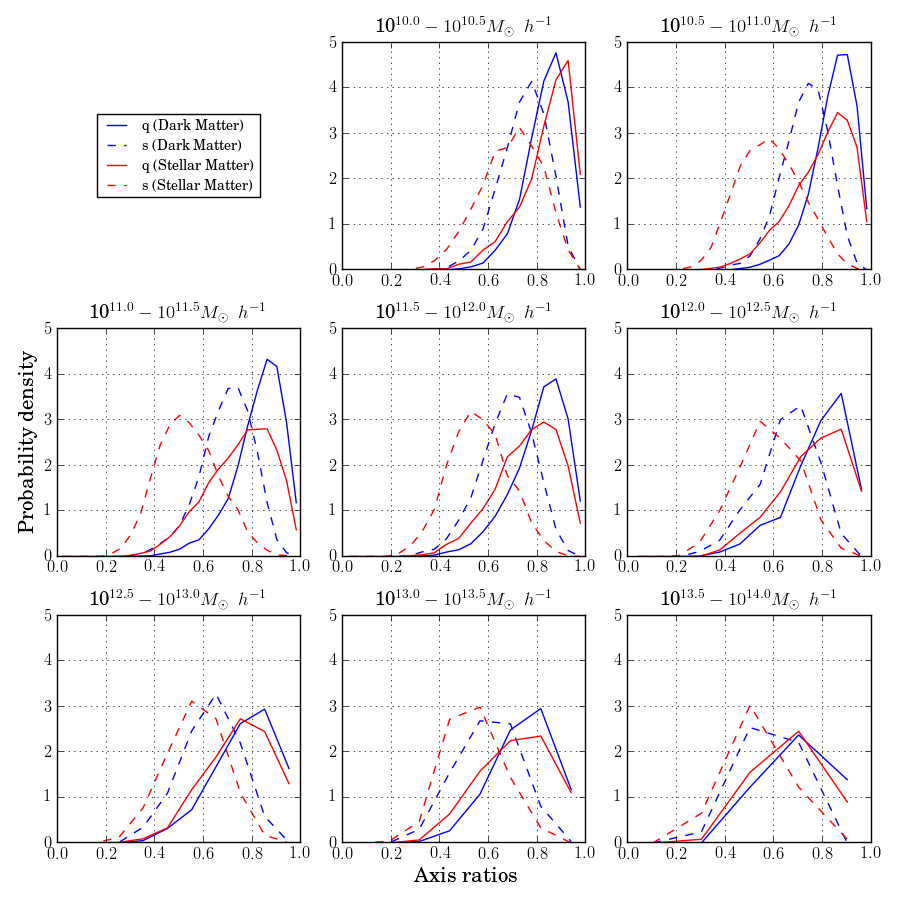}
\caption{\label{F:fig1}3d axis ratio distributions of dark matter and stellar matter in
  subhalos at $z = 0.06$, for masses of subhaloes in the range $10^{10.0}
  - 10^{14.0}$\hMsun.}
\end{center}
\end{figure*}
 
The distributions of axis ratios, $q~(b/a)$ and $s~(c/a)$ for dark matter and stellar
matter of subgroups at redshift $z = 0.06$ for different mass bins are shown in
Figure~\ref{F:fig1}. The plot shows that the axis ratios are larger for dark matter when
compared to stellar matter, indicating that the dark matter component of a subgroup is more
round than the stellar matter. Also, we observe that there is no significant evolution in the distribution of axis ratios in adjacent panels. We henceforth present our results in three mass bins : $10^{10.0} -
10^{11.5}$\hMsun, $10^{11.5} - 10^{13.0}$\hMsun, and $> 10^{13.0}$\hMsun. For convenience, we refer to these mass bins as $M1, M2$ and $M3$ respectively. In the mass bin $M3$, the largest subhalo mass is $1.4\times10^{14}$\hMsun\ at $z=1.0$, with a host halo mass of $1.6\times10^{14}$\hMsun; it grows to $6.0\times10^{14}$\hMsun\ at $z=0.06$ with a host halo mass of $7.2\times10^{14}$\hMsun. 

\subsection{Redshift evolution and mass dependence of 3D axis ratios}

\begin{figure*}
\begin{center}
$\begin{array}{c@{\hspace{0.5in}}c}
\includegraphics[width=3.2in,angle=0]{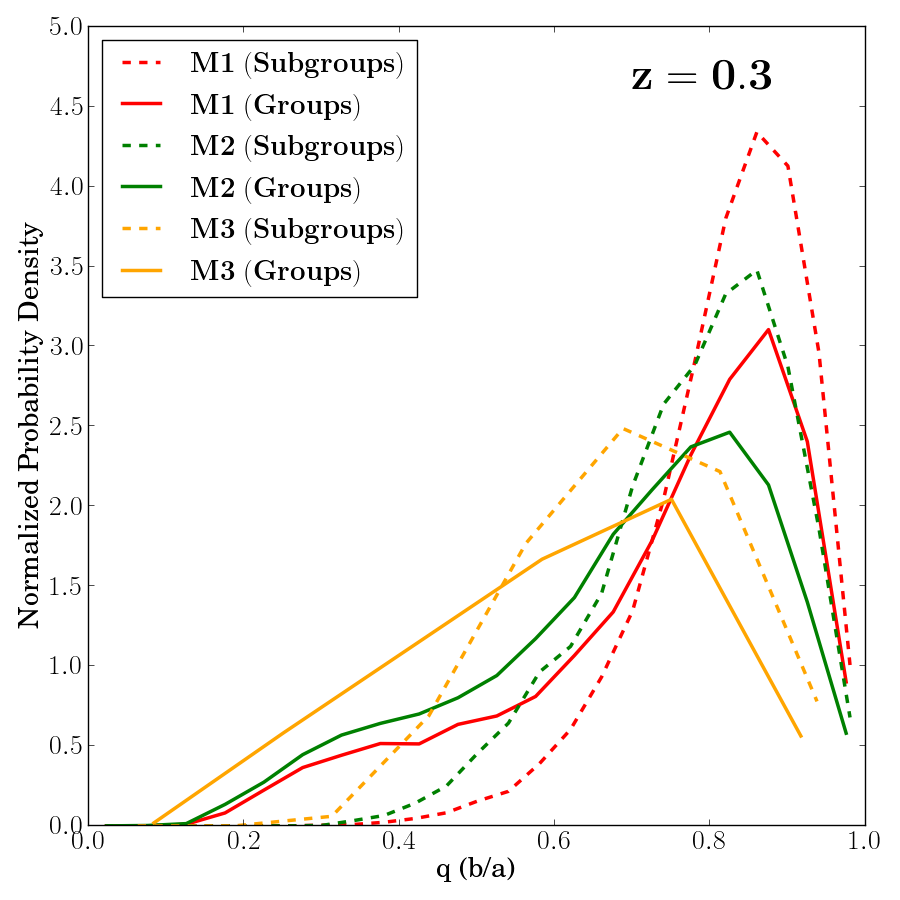} &
\includegraphics[width=3.2in,angle=0]{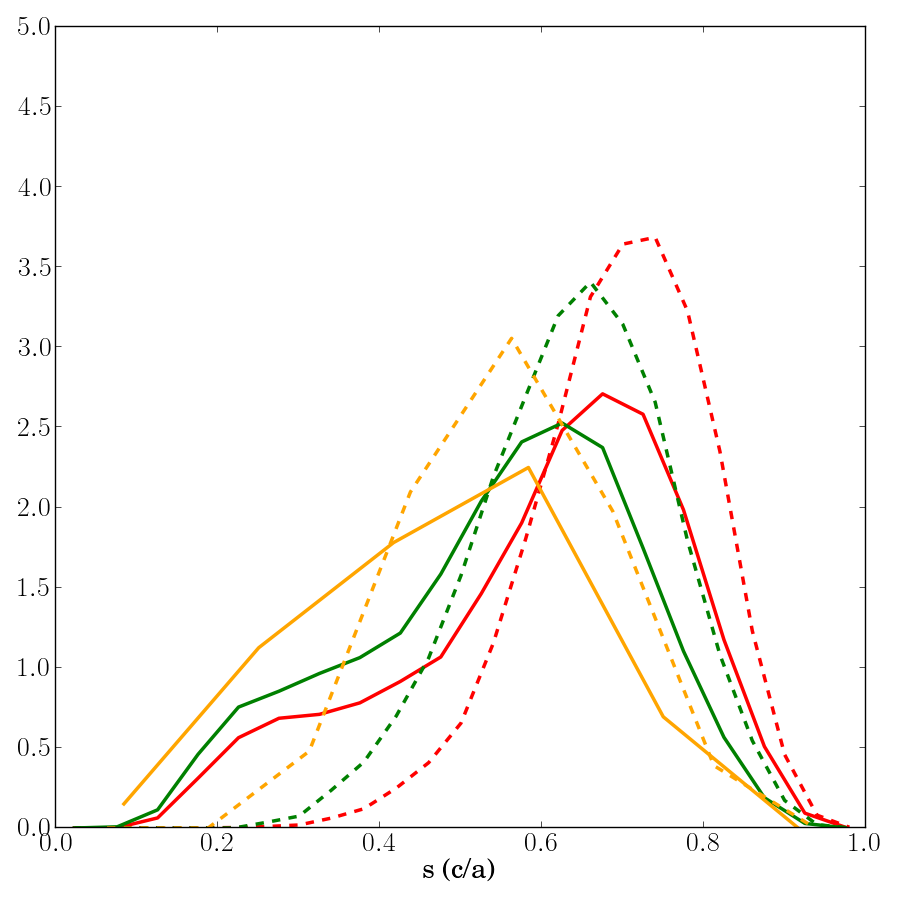}
\end{array}$
\caption{\label{F:fig3} Comparison of axis ratios, $q~(b/a)$ (left panel) and $s~(c/a)$ (right panel) between dark
  matter subgroups and groups at $z = 0.3$ in different mass bins.}
\end{center}     
\end{figure*}

\begin{figure*}
\begin{center}
$\begin{array}{c@{\hspace{0.5in}}c}
\includegraphics[width=3.2in,angle=0]{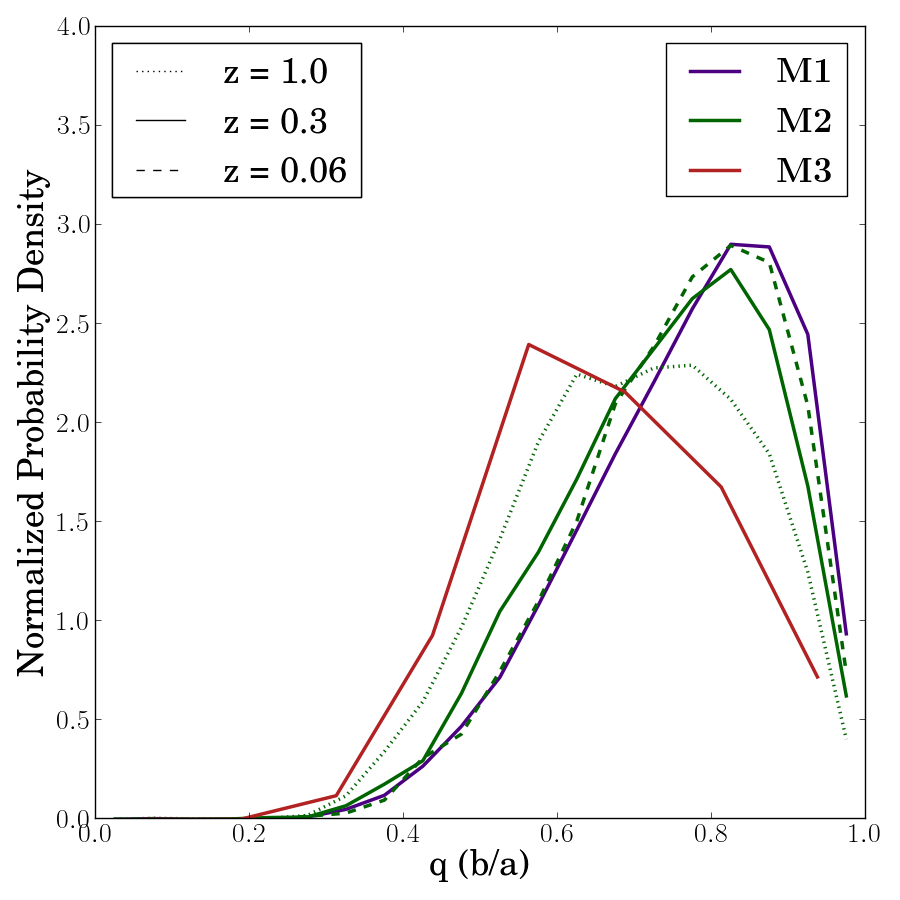} &
\includegraphics[width=3.2in,angle=0]{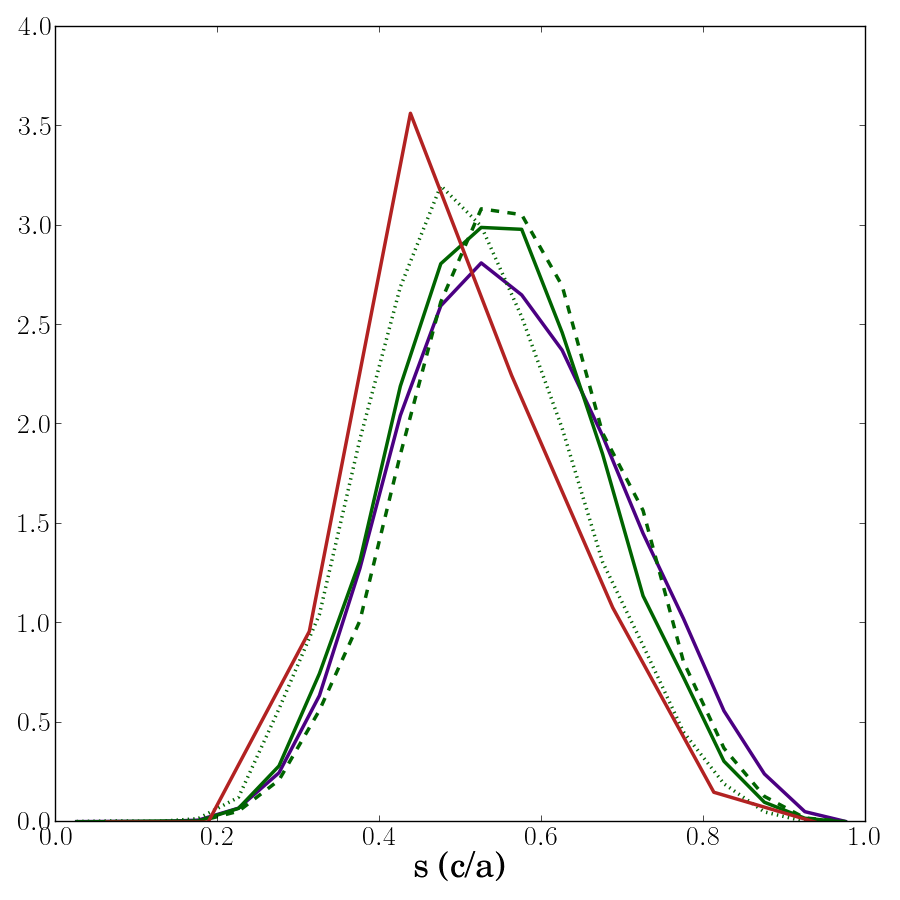}
\end{array}$
\caption{\label{F:fig4}Axis ratios $q~(b/a)$ (left panel) and $s~(c/a)$ (right panel) for stellar matter of subhalos
  at $z = 0.3$ in mass bins ($M1, M2$ and $M3$) and at $z = 1.0,
  0.06$ for the central mass bin, $M2$.}
\end{center}
\end{figure*}

\begin{figure*}
\begin{center}
$\begin{array}{c@{\hspace{0.5in}}c}
\includegraphics[width=3.2in,angle=0]{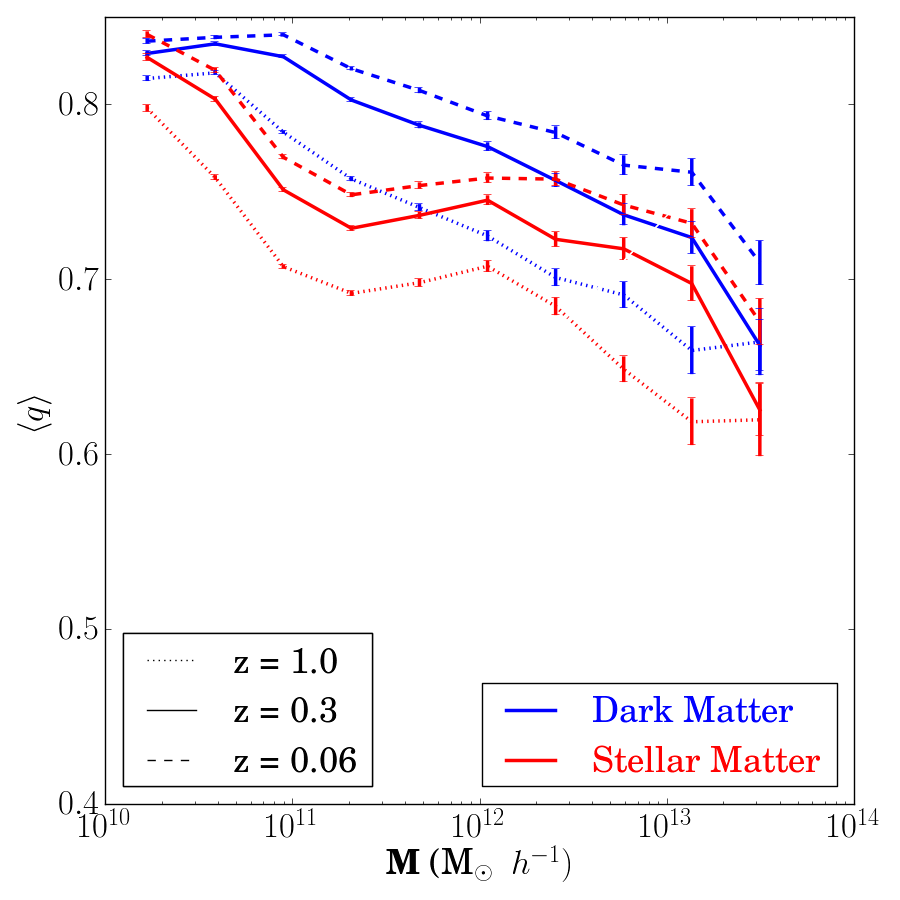} &
\includegraphics[width=3.2in,angle=0]{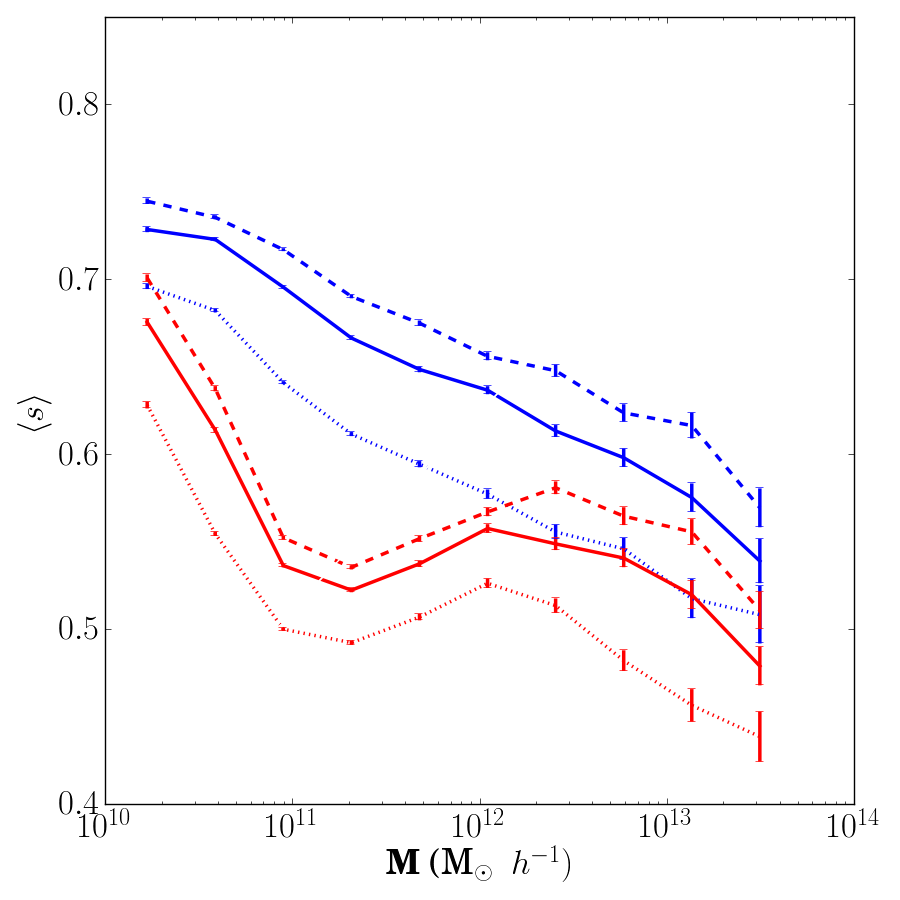}
\end{array}$
\caption{\label{F:fig_avgqs}Average axis ratios, $\langle q \rangle$ (left panel) and $\langle s
  \rangle$ (right panel) for dark matter and stellar component of subhalos as a function of mass, at
  redshifts $z = 1.0, 0.3$, and $0.06$.}
\end{center}
\end{figure*}

\begin{figure}
\begin{center}
\includegraphics[height=\columnwidth,angle=0]{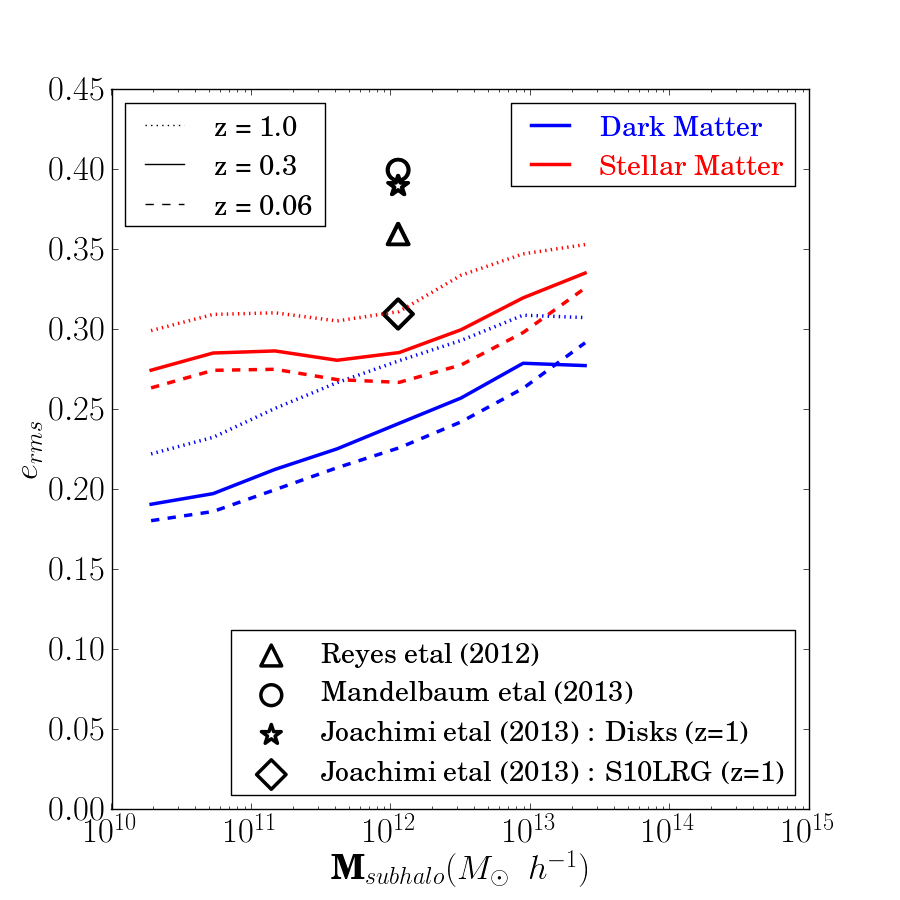}
\caption{\label{F:fig116}RMS ellipticity per component for projected shapes, $e_\text{rms}$, for dark matter and stellar matter at $z =
  1.0, 0.3$, and $0.06$
  as a function of cumulative subhalo mass.}
\end{center}
\end{figure}

In Figure~\ref{F:fig3}, we compare the distribution of axis ratios for groups and subgroups
at redshift $z = 0.3$ in different mass bins. Here, we consider the dark matter
component of groups and subgroups only. From the plot, we can see that for groups, as we
go to higher masses, the axis ratios decrease for both groups and subgroups. Comparing the
shape distributions between groups and subgroups, we can conclude that the shapes of
subgroups are more round when compared to groups in any given mass bin, in
agreement with the findings of \cite{2007ApJ...671.1135K} using dark matter-only
simulation. Even in hydrodynamic simulations, \cite{2006EAS....20...65K} found that dark
matter subhalos are more round than halos. We can also see that as we go to higher mass
bins, the axis ratios of dark matter halos and subhalos decrease in agreement with the
findings of \cite{2008MNRAS.388L..34K}.
  
To investigate the mass dependence of axis-ratio distributions for the
stellar matter component of subgroups, we plot the axis ratios $(q,s)$
at redshifts $z = 0.3$ in the mass bins $M1, M2$ and $M3$ in Fig.~\ref{F:fig4}. The plot shows that as we go
to higher mass bins, the shapes of subhalos get more flattened. Using
the distribution of satellites and Monte Carlo simulations,
\cite{2008MNRAS.385.1511W} reached the same conclusion for dark matter
halos. We find that the shapes of stellar matter also follow a similar
trend. To understand the redshift evolution of shapes, we also show
the shape distributions at $z = 1.0$, and $0.06$ for the middle mass
bin, $10^{11.5} - 10^{13.0}$\hMsun. The lines show that at lower
redshifts, the shapes tend to become
rounder. \cite{2005ApJ...618....1H}, \cite{2006MNRAS.367.1781A} and \cite{2012JCAP...05..030S} used $N$-body simulations and considered
the axis ratio distributions as a function of mass and redshift. Their
results show that at a given mass, halos are more round at lower
redshift, and more massive halos are more flattened which is
consistent with our findings. In Fig.~\ref{F:fig_avgqs}, we show the
average axis ratios, $\langle q \rangle$ and $\langle s \rangle$ as a
function of mass at different redshifts $z = 1.0, 0.3$, and $0.06$ for
the dark matter and stellar component. We also provide fitting
functions for the average axis ratios of the dark matter and stellar
component of subhalos as a function of mass and redshift in
Appendix~\ref{appa}. The plots for average axis ratios of the dark
matter component can be compared against
\cite{2006MNRAS.367.1781A}. Our results agree with theirs qualitatvely
in that the average axis ratios, $\langle q \rangle$ and $\langle s
\rangle$, increase as we go to lower redshifts and lower masses for
the dark matter component. Their curves show a lower average $\langle
s\rangle$ which may be because of the different criteria used in the
determination of halo shapes, changes in dark matter shapes from the
effect of baryons, and different cosmological parameters. Also, they
measured average axis ratios for halos, while our results are for
subhalos. For the stellar matter, we can see that in general, the
average axis ratios decrease with subhalo mass. However,  there is an
increase in the intermediate mass range around $\sim 10^{11}\hMsun$ followed by a decreasing trend
once again. We will investigate the dependence of this trend on the type and color of galaxies in a future study to understand the significance of this mass scale.

To compare the axis ratio distributions of projected shapes defined by stellar matter of subhalos with results from observational measurements, we use the statistic, rms ellipticity. The rms ellipticity per single component,  $e_\text{rms}$, is given by 
\begin{equation} \label{eq:meanrms}
 e_\text{rms}^2 = \frac{\sum_{i}{(\frac{1 - q_{i}'^2}{1 + q_{i}'^2})}^2}{2N},
\end{equation}
where $q_{i}' = \frac{b_{i}'}{a_{i}'}$ for the $i^{th}$ subgroup and
$N$ is the total number of subgroups considered. 
In Fig~\ref{F:fig116}, we show the projected rms ellipticity $e_\text{rms}$ as a function of cumulative
mass of subhalos (by considering all subhalos of mass greater a given mass) for redshifts $z = 1.0, 0.3$, and $0.06$. Our results can
be compared against those from observations in the Sloan Digital Sky
Survey (SDSS) given in \cite{2012MNRAS.425.2610R}. For stellar matter,
we obtained $e_\text{rms} = 0.28$ at $z = 0.3$ for $\msubhalo >
10^{12}$\hMsun, which is smaller than the observed value of $0.36$,
but reasonably close (and larger than that expected for dark matter
component). The catalogue used by \cite{2012MNRAS.425.2610R} has been corrected for measurement noise, but it has some selection effects that bias it
slightly in the direction of eliminating small round galaxies, thus
boosting the RMS ellipticity in the sample of galaxies selected in the
data compared to a fair sample. In addition to SDSS, we also made a comparison with observational results obtained using data from COSMOS survey. An $e_\text{rms} = 0.4$ is obtained using shapes from a galaxy sample in \cite{2013arXiv1308.4982M} with a median redshift of $\sim 0.67$. These galaxies correspond to a mass of $\sim 10^{12}\hMsun$ at the median redshift. We made further comparison with measurements on rms ellipticity presented in \cite{2013MNRAS.431..477J}. For a close comparison, we used the results presented for late-type disk dominated galaxies at $z=1$ with the  assumption that the sample of galaxies in the simulation is dominated by disks at this redshift. The observational measurements give an rms ellipticity per component of $\sim 0.39$ at $z=1$ which is higher than our values which are in the range of $0.3-0.35$. The lower rms ellipticity may be due to a lower fraction of disk-dominated galaxies in the simulations, or due to the disks not being perfectly realistic. Another comparison is made with a sample of elliptical red galaxies ($S10LRG$) given in \cite{2013MNRAS.431..477J} where the rms ellipticity per single component is measured to be $~\sim 0.31$ at $z = 1.0$, in agreement with our result, but no significant redshift evolution of $e_\text{rms}$ is detected for this sample. However, the fraction of galaxies in our simulated sample that are red is likely to be a function of redshift.  Also, in some of the observations \citep{2012MNRAS.425.2610R,2013MNRAS.431..477J}, the shape
estimator is weighted towards the inner part of the luminosity
distribution in a galaxy, while our shape measurements are obtained by
considering all the particles of a given type in the subhalo,
emphasizing the shape of stellar matter at large
radii (similar to the shape estimates in \citealt{2013arXiv1308.4982M} from fitting light profiles to galaxy models). Given the known differences between how the measurements in data and simulations were carried out, it is difficult to make a quantitative comparison, however, there are no red flags for a major discrepancy.


\section{Misalignments between stellar matter and dark matter shapes of
  subhalos}\label{S:misalignments}

In this section, we compare the major axis orientations of the stellar
components and dark matter components of subhalos, in 3D and 2D, in
order to quantify the degree of misalignment between them. We
investigate the dependence of the probability 
distribution of the misalignments on the mass range of subhalos and redshift. We also
discuss the change in misalignments in going from 3D, as defined by
the physics, to 2D, which is what we observe for real galaxies. Finally, the
misalignments are compared for centrals and satellite subgroups.

\subsection{Definition of misalignment angle}

For each subgroup, we determined the relative orientation of the major axis of its dark
matter subhalo with its stellar component. If $\hat{e}_{ga}$ and
$\hat{e}_{da}$ are the
major axes of the stellar and dark matter components, respectively, then the misalignment
angle is given by
\begin{equation} \label{eq:misalignangle}
 \theta_{m} = \arccos(\left|\hat{e}_{da} \cdot \hat{e}_{ga}\right|)
\end{equation}

The same definition can be used to determine the misalignment angle in 2D. It is to be noted here that the major axis is not well defined for ellipsoids which are nearly spherical. However, we verified that our results for misalignment angles do not change significantly when we exclude subhalos with $q$ and $s > 0.95$ for shapes defined by the dark matter or stellar matter.   
    
\subsection{Mass and redshift dependence of misalignments}

\begin{figure*}
\begin{center}
$\begin{array}{c@{\hspace{0.5in}}c}
\includegraphics[width=3.2in,angle=0]{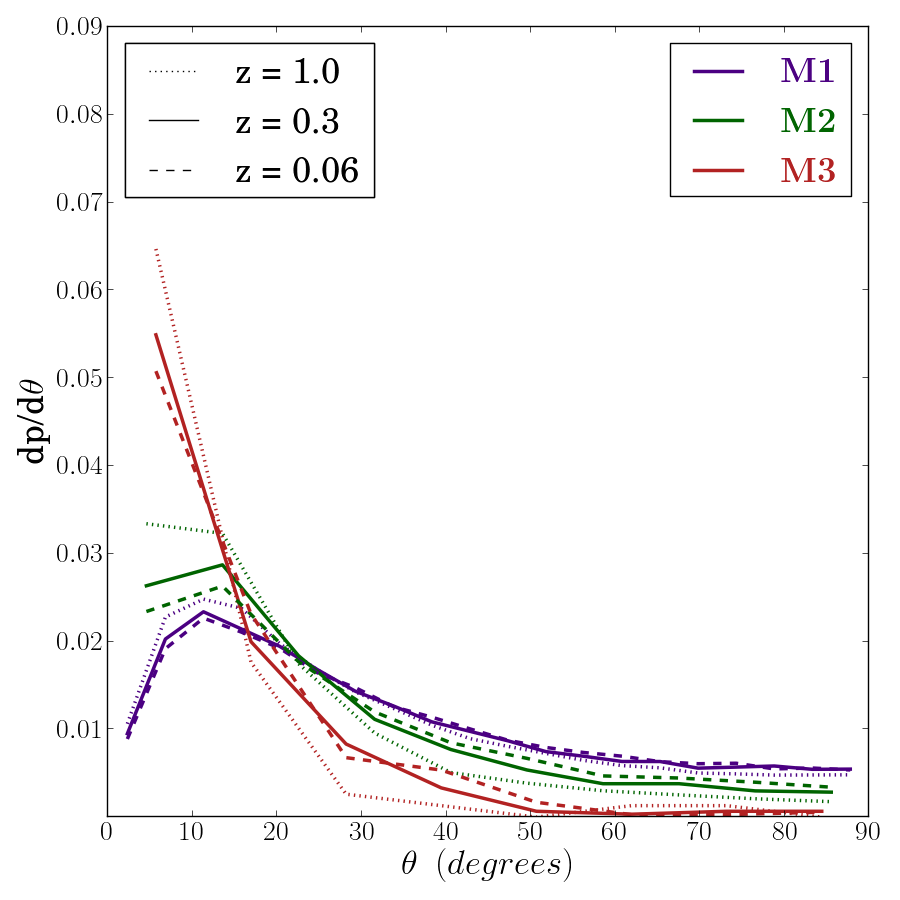} &
\includegraphics[width=3.2in,angle=0]{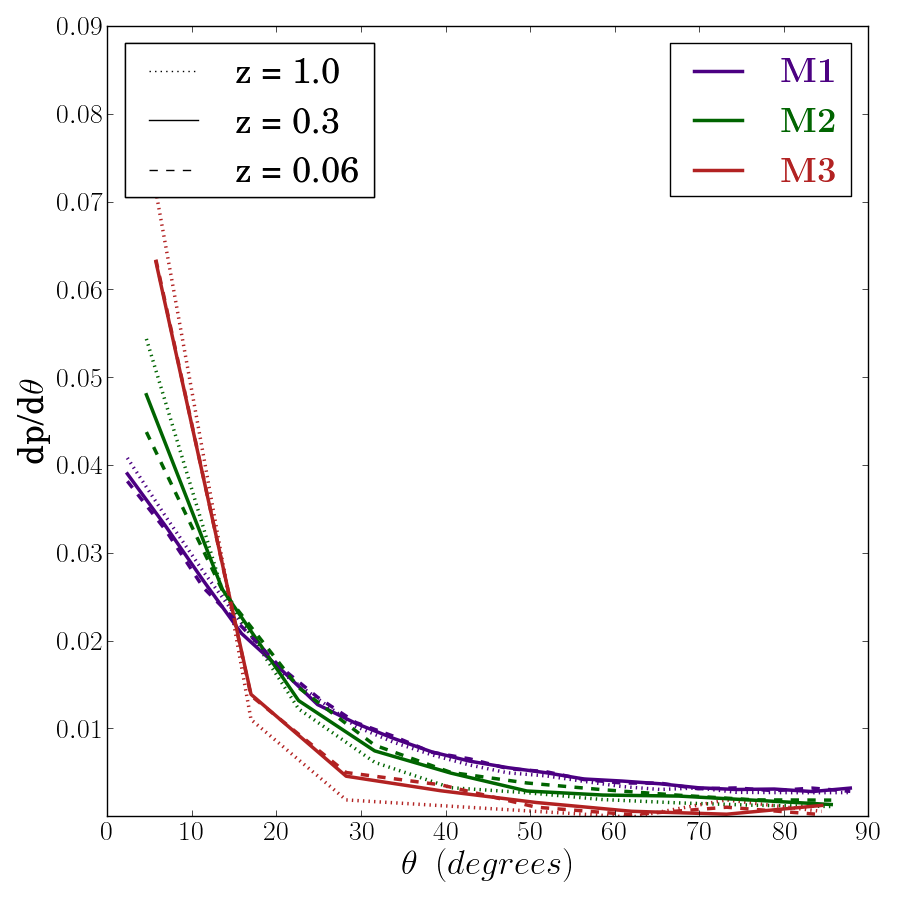}
\end{array}$
\caption{\label{F:fig6} Histogram of 3D (left panel) and 2D (right
  panel) misalignments at redshifts $z = 1.0, 0.3$, and 
  $0.06$ in the mass bins $M1, M2$ and $M3$.}
\end{center}
\end{figure*}

\begin{table}
\begin{center}
\caption{\label{T:tab1} Mean 3D misalignments in subgroups at
  redshifts $z = 1.0, 0.3$, and
  $0.06$ in the mass bins $M1, M2$ and $M3$.}
\begin{tabular}{@{}lccc}
\hline
Mass (\hMsun) & \multicolumn{3}{c}{Mean 3D misalignment angle}\\
 & $z = 1.0$ & $z = 0.3$ & $z = 0.06$ \\
\hline
$M1: 10^{10.0}-10^{11.5}$ & $31.61^{\circ}$ & $33.47^{\circ}$ & $34.10^{\circ}$\\
$M2: 10^{11.5}-10^{13.0}$ & $20.98^{\circ}$ & $25.20^{\circ}$ & $27.73^{\circ}$\\
$M3: ~> 10^{13.0}$ & $10.00^{\circ}$ & $13.04^{\circ}$ & $13.87^{\circ}$\\
\end{tabular}
\end{center}
\end{table}

\begin{table}
\begin{center}
\caption{\label{T:tab2} Mean 2D misalignments in subgroups at
  redshifts $z = 1.0, 0.3$, and 
  $0.06$ in mass bins $M1, M2$ and $M3$}
\begin{tabular}{@{}lccc}
\hline
Mass (\hMsun) & \multicolumn{3}{c}{Mean 2D misalignment angle}\\
 & $z = 1.0$ & $z = 0.3$ & $z = 0.06$ \\
\hline
$M1: 10^{10.0}-10^{11.5}$ & $22.61^{\circ}$ & $23.78^{\circ}$ & $23.88^{\circ}$\\
$M2: 10^{11.5}-10^{13.0}$ & $15.51^{\circ}$ & $17.89^{\circ}$ & $19.41^{\circ}$\\
$M3: ~> 10^{13.0}$ & $8.74^{\circ}$ & $10.73^{\circ}$ & $11.00^{\circ}$\\
\end{tabular}
\end{center}
\end{table}

In Fig.~\ref{F:fig6}, we show the misalignment probability
distributions for subgroups at redshifts $z = 1.0,0.3$, and $0.06$ in
mass bins $M1, M2$ and $M3$. From the plots,
we see that in the massive bins, the stellar component is more
strongly aligned with its dark matter subhalos. The mean 3D
misalignments for each mass bin are listed in Table~\ref{T:tab1}. As
we go from lower to higher mass bins, the mean misalignments decrease
from $34.10^{\circ}$ to $13.87^{\circ}$. For a given mass bin, the
misalignment strength increases towards lower redshifts, as shown in the
plot and table; however, the trend with mass is far stronger than the
trend with redshift. When comparing 3D and 2D misalignments, we find
that the misalignments are more prominent in the 3D situation. This is mainly due to a decrease in misalignment angle by projecting along a particular direction. Also, if we consider random distribution of misalignment angles, it can be inferred geometrically that the probability increases with angle of misalignment in 3D, while the distribution is uniform in 2D. In
Appendix~\ref{appb}, we give fitting functions for the probability
distributions of 3D and 2D misalignment angles in different mass bins
at redshifts $z = 1.0, 0.3, 0.06$. These probability distributions of misalignment angles are useful in predicting intrinsic alignment signals and estimating the $C_{1}$ parameter (overall alignment strength) in the linear alignment model \citep{2011JCAP...05..010B}. Table~\ref{T:tab2} shows the mean
misalignments in 2D.  The fitting functions for mean misalignment
angles as a function of mass are given in Appendix~\ref{appc}. The
misalignment distribution for masses $\msubhalo > 10^{13}$\hMsun\
shows that the stellar shapes are well aligned with their host dark
matter subhalos with a mean misalignment angle of $10.00^{\circ}$ at
$z = 1$ and $13.87^{\circ}$ at $z = 0.06$. In a similar mass range,
using $N$-body simulations, \cite{2009ApJ...694..214O} assumed a
gaussian distribution of misalignment angle with zero mean and
constrained the width, $\sigma_{\theta}$, to be around $35^{\circ}$ so
as to match the observed ellipticity correlation functions for central
LRGs. This corresponds to an absolute mean misalignment angle of $\sim28^{\circ}$. The galaxies used by \cite{2009ApJ...694..214O} have masses corresponding to our
highest mass bin, for which we predict a stronger alignment between
dark matter halo and galaxy; however, because of the different
methodology used to indirectly derive their misalignment angle
compared to our direct prediction from simulations, a detailed
comparison is difficult.

\section{Shape distributions and misalignments for central
  vs. satellite galaxies}\label{S:censat}

Here we consider the axis ratio distributions and misalignment probability distributions
for central and satellite subgroups in different mass bins, divided in
two ways: based on the parent halo
mass and based on the individual subhalo mass.

\begin{figure*}
\begin{center}
$\begin{array}{c@{\hspace{0.5in}}c}
\includegraphics[width=3.2in,angle=0]{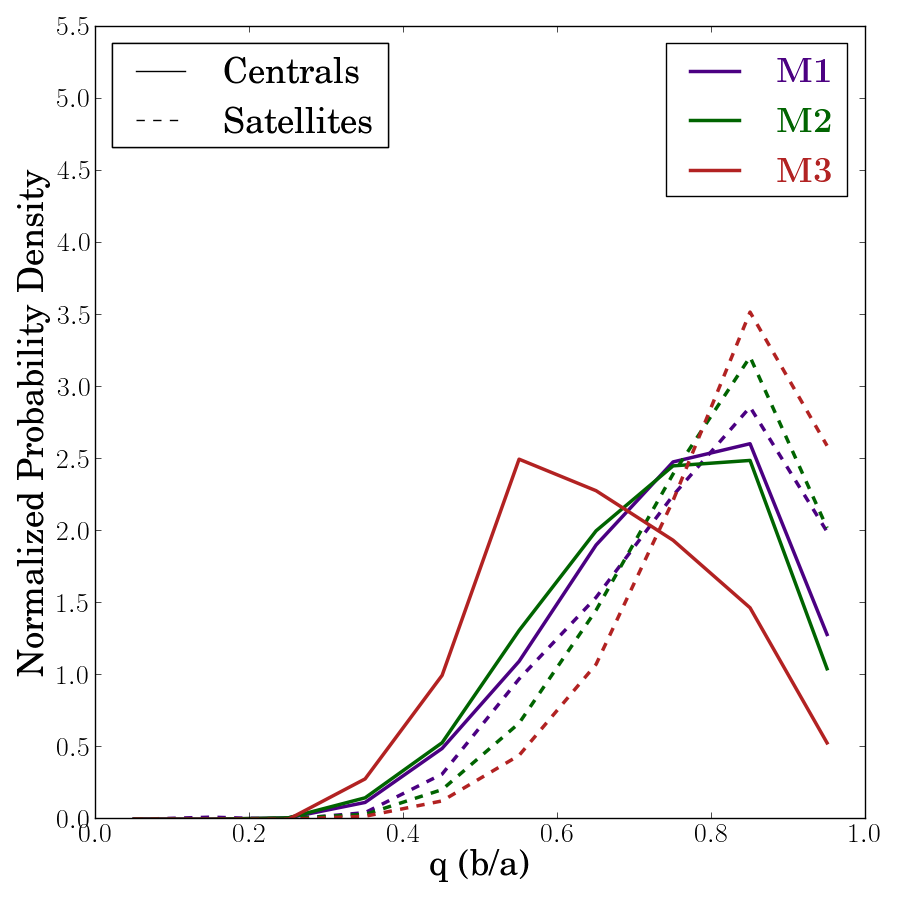} &
\includegraphics[width=3.2in,angle=0]{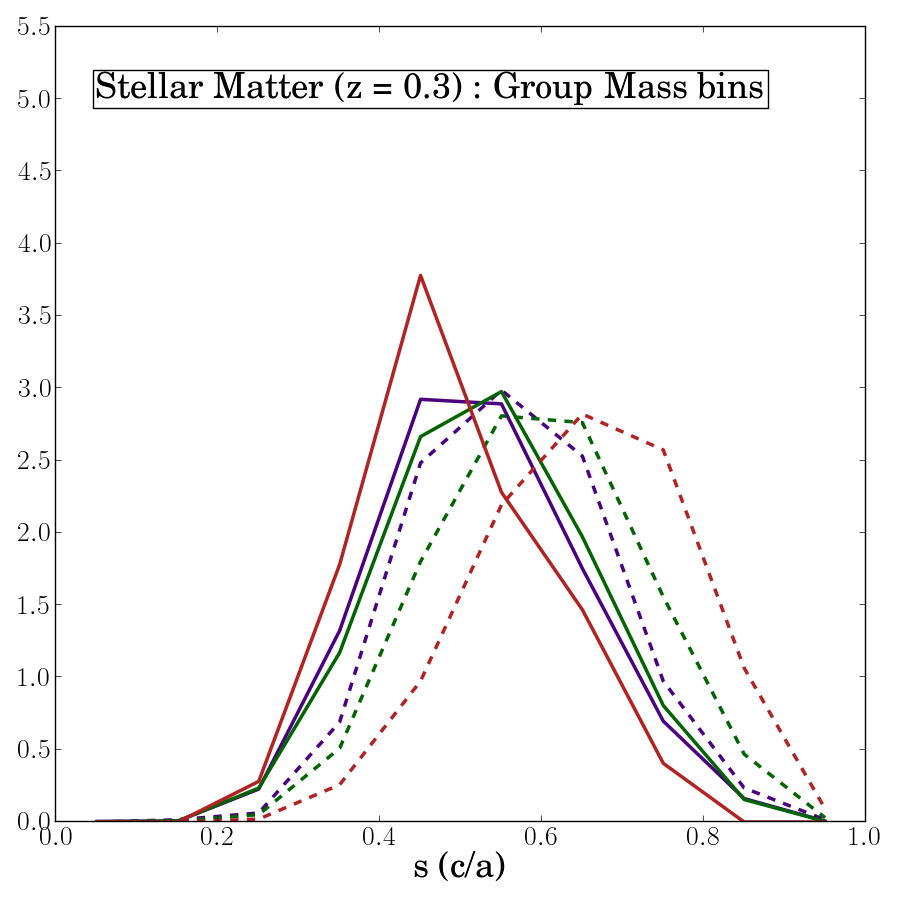}\\
\includegraphics[width=3.2in,angle=0]{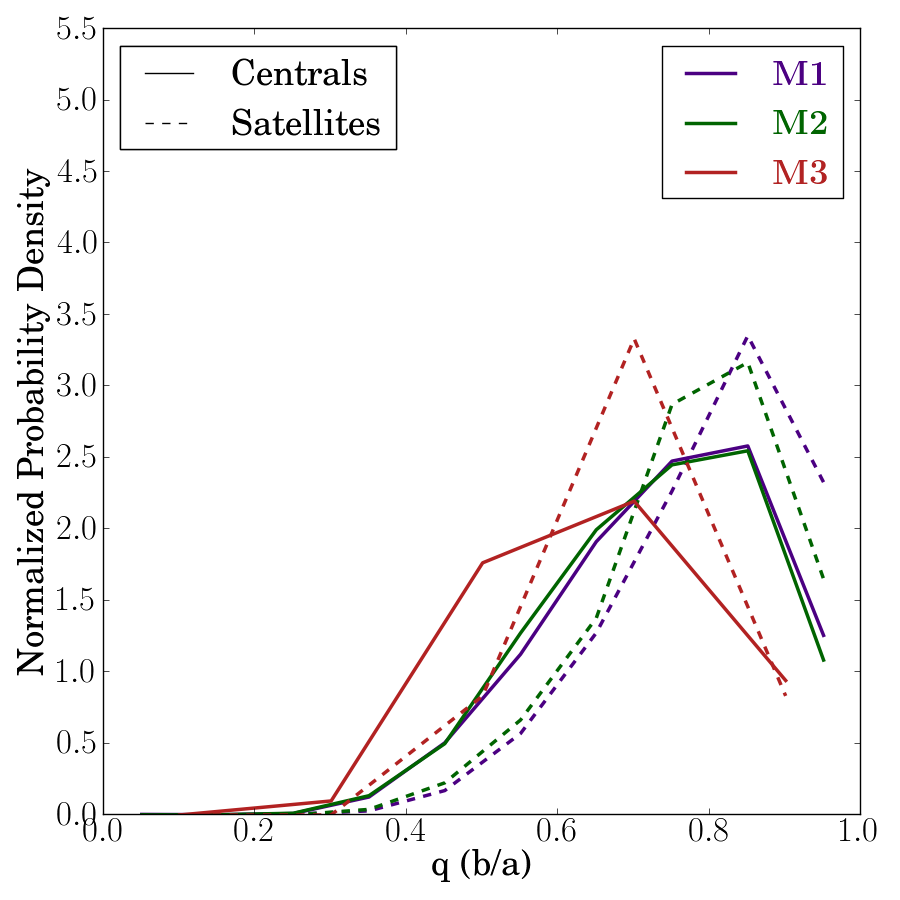} &
\includegraphics[width=3.2in,angle=0]{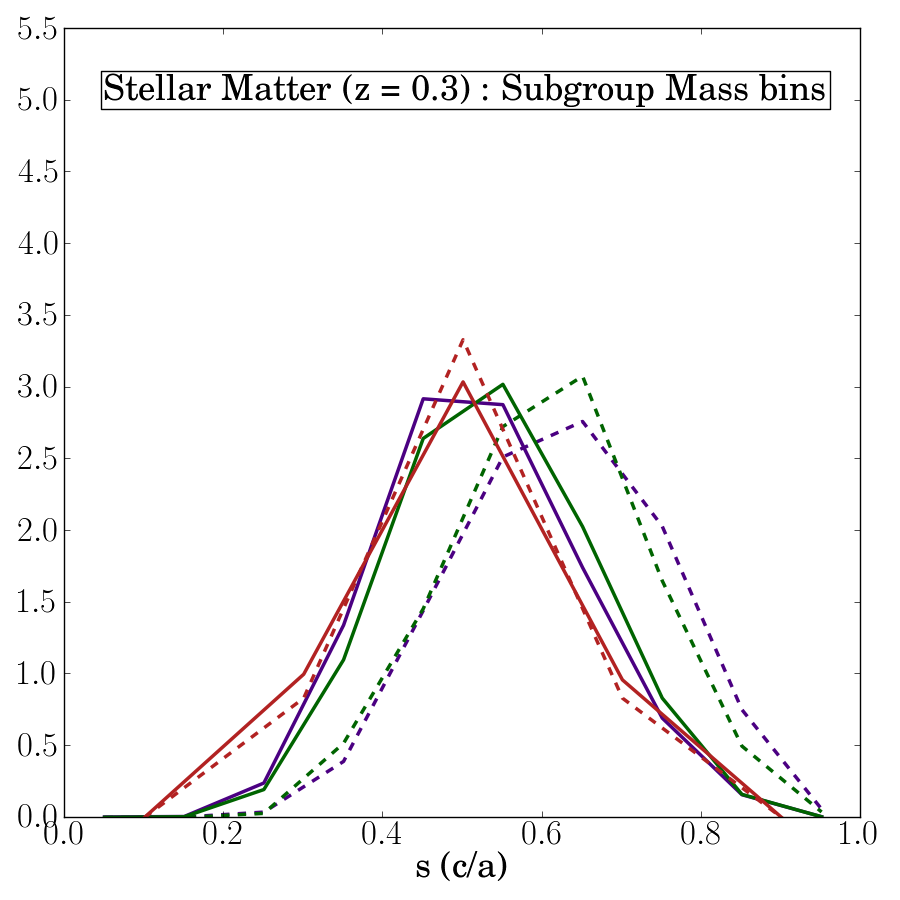}\\
\end{array}$
\caption{\label{F:fig_avgqs_censat}Axis ratio distributions of stellar matter in subgroups
  for centrals and satellites in mass bins $M1, M2$ and $M3$. {\em
    Top panel:} Results when dividing based on the parent halo mass;
  {\em bottom panel:} when dividing based on the subhalo mass.  In
  both rows, the left and right panels show results for $q$ and $s$, respectively.}
\end{center}
\end{figure*}

\begin{figure*}
\begin{center}
$\begin{array}{c@{\hspace{0.5in}}c}
\includegraphics[width=3.2in,angle=0]{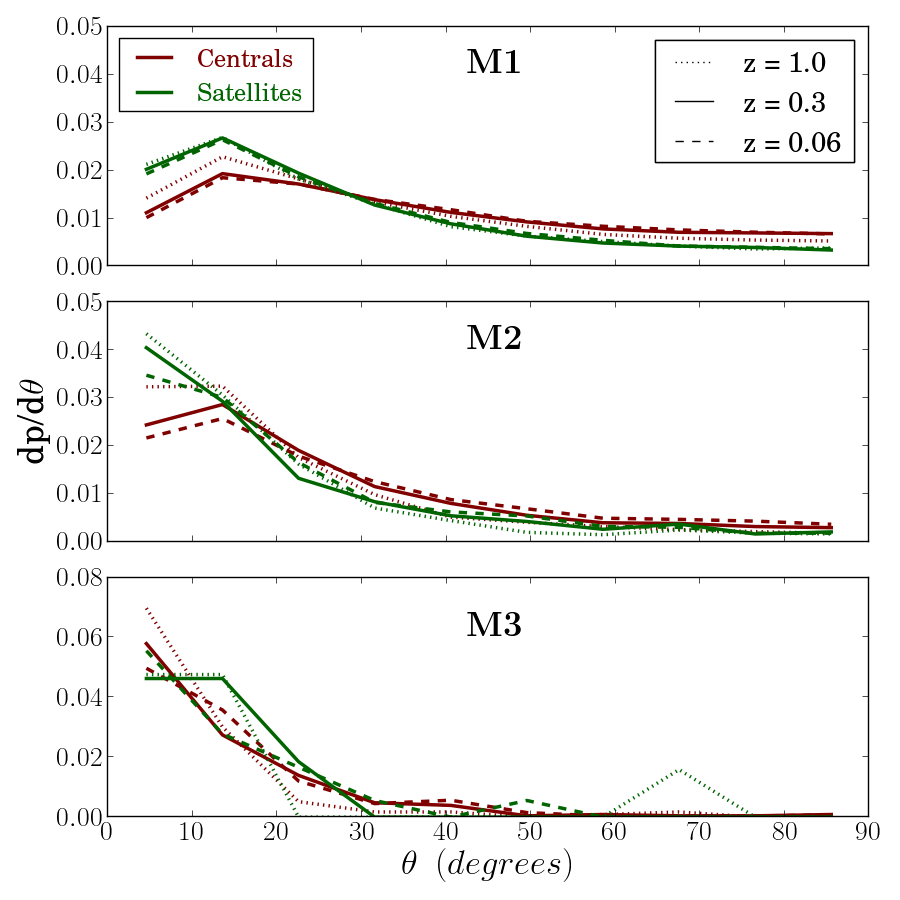} &
\includegraphics[width=3.2in,angle=0]{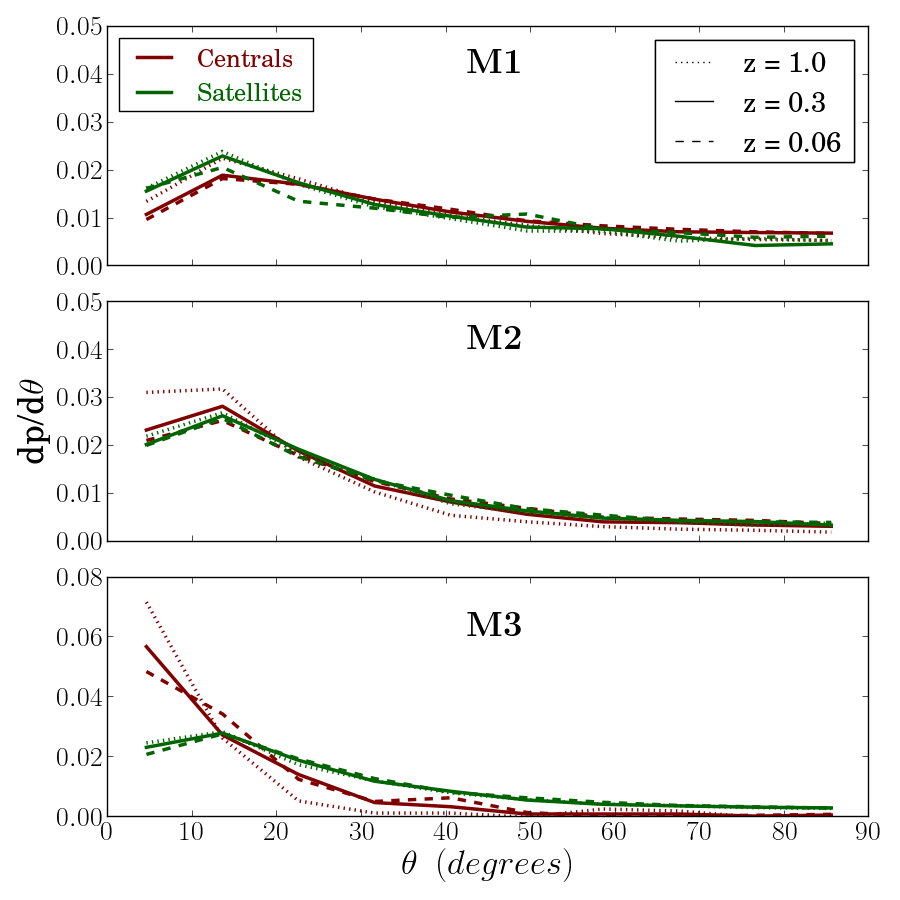}
\end{array}$
\caption{\label{F:fig15}Histograms of misalignment angles for central and satellite
  subgroups in mass bins $M1, M2$ and $M3$. {\em Left:} Results when dividing
  based on the subhalo mass; {\em right:} when dividing based on the 
  halo mass.}
\end{center}
\end{figure*}
 
\begin{table*}
\begin{center}
\caption{\label{T:tab3} Mean 3D misalignments in central and satellite subgroups at
  redshifts $z = 1.0, 0.3$, and $0.06$ in subhalo mass bins $M1, M2$ and $M3$.}
\begin{tabular}{@{}lcccccc}
\hline
 & \multicolumn{2}{c}{$z = 1.0$} & \multicolumn{2}{c}{$z = 0.3$} & \multicolumn{2}{c}{$z = 0.06$} \\
\hline
Subhalo Mass (\hMsun) & Centrals & Satellites & Centrals & Satellites & Centrals & Satellites \\
\hline
$M1: 10^{10.0}-10^{11.5}$ & $33.42^{\circ}$ & $28.21^{\circ}$ & $37.07^{\circ}$ & $28.22^{\circ}$ & $37.83^{\circ}$ & $29.00^{\circ}$\\
$M2: 10^{11.5}-10^{13.0}$ & $21.30^{\circ}$ & $18.03^{\circ}$ & $25.85^{\circ}$ & $20.43^{\circ}$ & $28.68^{\circ}$ & $21.54^{\circ}$\\
$M3: ~> 10^{13.0}$ & $9.61^{\circ}$ & $17.17^{\circ}$ & $13.11^{\circ}$ & $11.73^{\circ}$ & $14.00^{\circ}$ & $12.03^{\circ}$\\
\end{tabular}
\end{center}
\end{table*}

\begin{table*}
\begin{center}
\caption{\label{T:tab4} Mean 3D misalignments in central and satellite subgroups at
  redshifts $z = 1.0, 0.3$, and $0.06$ in parent halo mass bins $M1, M2$ and $M3$.}
\begin{tabular}{@{}lcccccc}
\hline
 & \multicolumn{2}{c}{$z = 1.0$} & \multicolumn{2}{c}{$z = 0.3$} & \multicolumn{2}{c}{$z = 0.06$} \\
\hline
Halo Mass (\hMsun) & Centrals & Satellites & Centrals & Satellites & Centrals & Satellites \\
\hline
$M1: 10^{10.0}-10^{11.5}$ & $33.88^{\circ}$ & $32.88^{\circ}$ & $37.39^{\circ}$ & $32.71^{\circ}$ & $38.12^{\circ}$ & $35.60^{\circ}$\\
$M2: 10^{11.5}-10^{13.0}$ & $21.98^{\circ}$ & $27.76^{\circ}$ & $26.61^{\circ}$ & $28.52^{\circ}$ & $29.10^{\circ}$ & $29.32^{\circ}$\\
$M3: ~> 10^{13.0}$ & $10.33^{\circ}$ & $26.10^{\circ}$ & $13.47^{\circ}$ & $26.48^{\circ}$ & $14.76^{\circ}$ & $27.36^{\circ}$\\
\end{tabular}
\end{center}
\end{table*}
In the top panel of Fig.~\ref{F:fig_avgqs_censat}, we show normalized histograms of
$q$ and $s$ for centrals and satellites binned according
to their parent halo mass, for the bins, $M1, M2$ and $M3$. In the bottom panel,
we show the same thing, but dividing based on the subgroup masses. The plots show
that satellite subgroups are more round than central subgroups. 
For satellites, we see that the axis ratio distributions show a strong
dependence on the subhalo mass and, for $s$, the parent halo mass.
These trends go in the opposite direction: satellites tend to have a
lower value of $s$ when their parent halo mass is low, or when their
subhalo mass is high. 
If we compare the top and bottom right figures, the
minor-to-major axis ratio distributions for centrals exhibit little
mass dependence when binning by subhalo mass, but more mass dependence
when binning by parent halo mass, suggesting an interesting
environment dependence.  

In Fig.~\ref{F:fig15}, we show the distributions of the misalignment
angles for central and satellite subgroups in different mass bins at
redshifts $z=1.0,0.3$, and $0.06$. In the right panel, the binning is
based on halo mass, while in the left panel, the binning is according
to subhalo mass. We can see that both centrals and satellites exhibit
the same qualitative features in the distributions of misalignment
angles as the whole sample of subgoups in
Fig.~\ref{F:fig6}. Tables~\ref{T:tab3} and~\ref{T:tab4} show the mean
misalignment angles of centrals and satellites binned binned according
to their subhalo and parent halo masses, respectively. Considering
mass bins based on individual masses of subhalos, we see that in
general, the degree of alignment is larger for satellites than for
centrals for all mass bins. However, if we bin based on the mass of
the parent halo, then at higher halo masses, central subgroups tend to
have larger alignments than the satellite subgroups. This effect may
be due to the centrals having higher masses than the satellites, which
tends to correlate with having a higher degree of alignment.
 
\section{Conclusions}\label{S:conclusions}

In this study, we used the MBII cosmological hydrodynamic simulation to study halo and galaxy shapes
and alignments, which are relevant for determining the intrinsic
alignments of galaxies, an important contaminant for weak lensing
measurements with upcoming large sky surveys.
While $N$-body simulations have been used in the past to study
intrinsic alignments, it is also important to study the effects due to
inclusion of the physics of galaxy formation; this includes effects
both on the overall shapes (ellipticities) of the galaxies and halos,
but also on any misalignment between them.  In order to study this
particular issue, we measured the shapes of dark matter and stellar
component of groups and subgroups.

Previous studies have used $N$-body simulations to study the mass
dependence and redshift evolution of the shapes of dark matter halos
and subhalos
\citep{{2005ApJ...632..706L},{2006MNRAS.367.1781A},{2007ApJ...671.1135K},{2008MNRAS.385.1511W},{2008MNRAS.388L..34K},{2012JCAP...05..030S}}. Our
results are qualitatively consistent with several trends identified in
previous work.  The first such trend that we confirm using SPH
simulations is that subhalos are more round than halos
\citep{{2006EAS....20...65K},{2007ApJ...671.1135K}}. The second trend
that we confirm is that the shapes of less massive subhalos are more
round than more massive subhalos
\citep{{2008MNRAS.388L..34K},{2008MNRAS.385.1511W}} and as we go to
lower redshifts, the subhalos also tend to become rounder
\citep{{2005ApJ...618....1H},{2006MNRAS.367.1781A},{2012JCAP...05..030S}}.

The effect of including baryonic physics on the shapes of dark matter
halos was studied previously using hydrodynamic simulations in a box
of smaller size and resolution compared to ours
\citep{{2006EAS....20...65K},{2010MNRAS.405.1119K},{2013MNRAS.429.3316B}}. \cite{2006EAS....20...65K}
found that the axis ratios of dark matter halos increase due to the
inclusion of gas cooling, star formation, metal enrichment, thermal
supernovae feedback and UV heating. \cite{2013MNRAS.429.3316B} found
that there is no major effect on shapes under strong feedback, but
they observed a significant change in the inner halo shape
distributions. \cite{2010MNRAS.405.1119K} found that there is no
effect on the shapes of dark matter subhaloes, where they included gas
dynamics, cooling, star formation and supernovae feedback.  Here, we
took advantage of the extremely high resolution of MBII to directly
study the mass dependence and redshift evolution of the shapes of the
stellar component of subhalos in addition to dark matter. However, we
did not study the effect of baryonic physics on dark matter shapes by
comparison with a reference dark matter only simulation in this work.

We found that the shapes of the dark matter component of subhalos are
more round than the stellar component. Similar to dark matter subhalo
shapes, the shapes of the stellar component also become more round as we
go to lower masses of subhalos and lower redshifts. We are also able
to calculate the projected rms ellipticity per single component for
stellar matter of subhalos, which can be directly compared with
observational results in \cite{2012MNRAS.425.2610R}. While the
observed result is 0.36 at the given mass range, from our simulation,
we measured a value of 0.28 at $z = 0.3$ for $M > 10^{12}$\hMsun,
which is close, particularly given the uncertainties that result from
observational selection effects that are not present in the
simulations and that drive the RMS ellipticity to larger values, and
given the different radial weighting in the two measurements.

By modelling subhalos as ellipsoids in 3D, we are able to calculate
the misalignment angle between the orientation of dark matter and
stellar component. Previous studies of misalignments in simulations
used either low-resolution hydrodynamic simulations, or $N$-body
simulations with a scheme to populate halos with galaxies and assign a
stochastic misalignment angle and other assumptions
\citep{{2005ApJ...628...21S},{2006MNRAS.371..750H},{2009RAA.....9...41F},{2009ApJ...694..214O},{2010MNRAS.405..274H},{2011MNRAS.415.2607D}}. By
direct calculation from our high-resolution simulation data, we found
that in massive subhalos, the stellar component is more aligned with
that of dark matter, qualitatively similar to results that have been
inferred previously through other means. For instance, at $z = 0.06$,
the mean misalignment angles in mass bins from $10^{10.0} -
10^{11.5}$\hMsun, $10^{11.5} - 10^{13.0}$\hMsun, and $10^{13.0} -
10^{15.0}$\hMsun\ are $34.10^{\circ}$, $27.73^{\circ}$,
$13.87^{\circ}$, respectively. The amplitude of misalignment increases
as we go to lower redshifts. The total mean misalignment angle of
$30.05^{\circ}$, $30.86^{\circ}$, $32.71^{\circ}$ at $z = 1.0$, $0.3$,
$0.06$ respectively shows an increasing trend, though the trend is far
weaker than trends with mass at fixed redshift. We also found that the
misalignments are larger for 3D shapes when compared to projected
shapes. It is to be noted here that we have not split our sample of subhalos according to the type of galaxy. The dependence of our results on galaxy type or color will be investigated in a future study. It is fairly well established that the alignment mechanism for discs and elliptical galaxies is different, so this is a necessary next step to obtain measurements which can be directly compared against observations and used as input for modeling.

Finally, we considered the axis ratios and misalignments in central
and satellite subgroups according to their parent halo mass and
individual mass of subgroups. We concluded that the shape of stellar component of satellites is more round than that of centrals. We also conclude that the satellite subgroups are
more aligned when compared to centrals in similar mass range. Observationally, it is not possible to
directly measure the misalignments in centrals and
satellites. Misalignment studies for central galaxies were done
earlier by \cite{{2008MNRAS.385.1511W},{2009ApJ...694..214O}}. Using
data and Monte Carlo simulations, \cite{2008MNRAS.385.1511W} predict a
Gaussian distribution of misalignment angle with zero mean and a
standard deviation of $23^{\circ}$ for their sample of red and blue
centrals. \cite{2009ApJ...694..214O} used $N$-body simulations and an
HOD model for assigning galaxies to halos. The alignment of central
LRG's with host halos is assumed to follow a Gaussian distribution
with zero mean. Okumura et al. arrived at a standard deviation of $35^{\circ}$
to match the observed ellipticity correlation. Our predictions of
misalignments for central and satellite subgroups are direct measurements that could be done through hydrodynamic simulations which include the physics of star formation.

In conclusion, we found that the axis ratios of the shapes of stellar
component of subhalos are smaller when compared to that of dark
matter. The shapes of both dark matter and stellar component tend to
become more round at low masses and low redshifts. We measured the
misalignment between the shapes of dark matter and stellar component
and found that the misalignment angles are larger at lower masses and
increase slightly towards lower redshifts. We found that the dependence is
more on the mass of subhalo than redshift. Finally, we split our
subhalos sample into centrals and satellites and found that in similar
mass range, the satellites have smaller misalignment angles.

We initiated this study with the goal of predicting intrinsic
alignments and constraining their impact on weak gravitational lensing
measurements. In this paper, we presented our results on the axis
ratios and orientations of both the dark matter and stellar matter of
subhalos. Future work will include the dependence of these results on
the radial weighting function used to measure the inertia tensor (as
in \citealt{2012JCAP...05..030S}), galaxy type and the difference between the
shape of the stellar mass versus of the luminosity distribution. We
will also present our results on the intrinsic alignment two-point correlation
functions in a future paper. Finally, future work should include
investigation of the impact of changes in the prescription for
including baryonic physics in the simulations. 

\section*{Acknowledgments}

RM's work on this project is supported in part by the Alfred P. Sloan
Foundation.  We thank Alina Kiessling, Michael Schneider, and Jonathan
Blazek for useful discussions of this work.
The simulations were run on the Cray XT5 supercomputer
Kraken at the National Institute for Computational Sciences. This
research has been funded by the National Science Foundation (NSF)
PetaApps programme, OCI-0749212 and by NSF AST-1009781.

\bibliographystyle{mn2e}
\bibliography{psubv1.bib}

\appendix
\section{Functional forms for dark matter and stellar matter shapes}
\label{appa}

\begin{figure*}
\begin{center}
$\begin{array}{c@{\hspace{0.5in}}c}
\includegraphics[width=3.2in,angle=0]{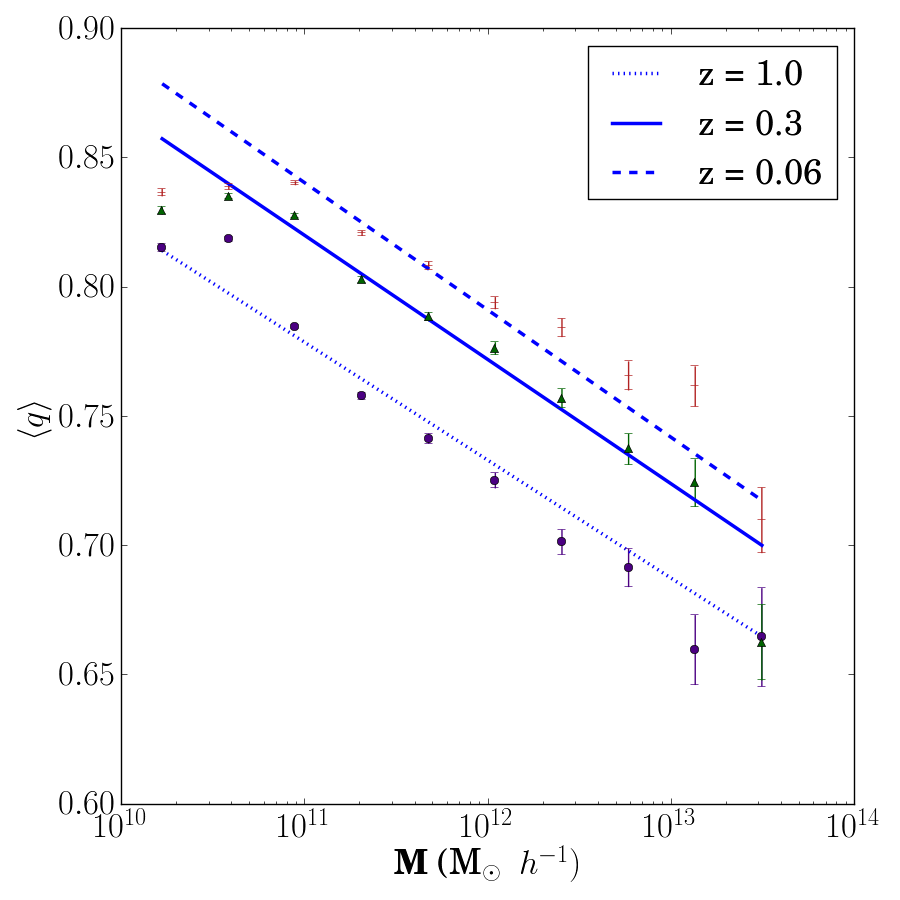} &
\includegraphics[width=3.2in,angle=0]{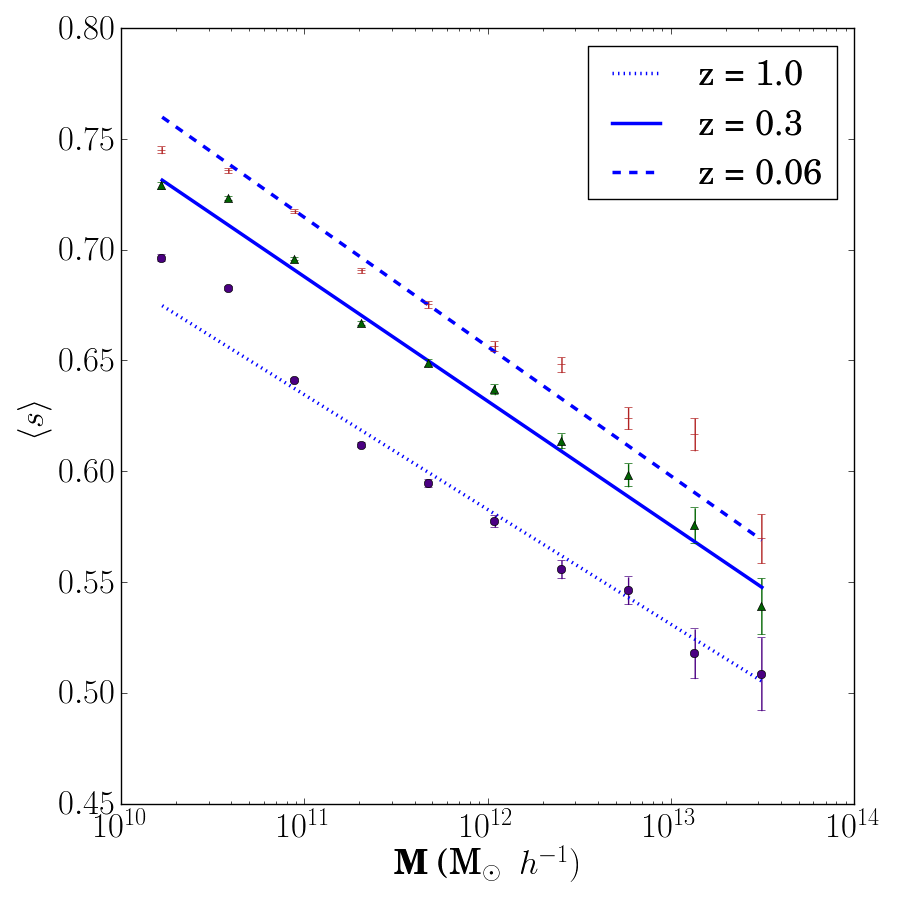}
\end{array}$
\caption{\label{F:fig_fit1}Fits for the axis ratios of shape defined by dark matter in subhalos as a function of mass and redshift}
\end{center}
\end{figure*}

\begin{figure*}
\begin{center}
$\begin{array}{c@{\hspace{0.5in}}c}
\includegraphics[width=3.2in,angle=0]{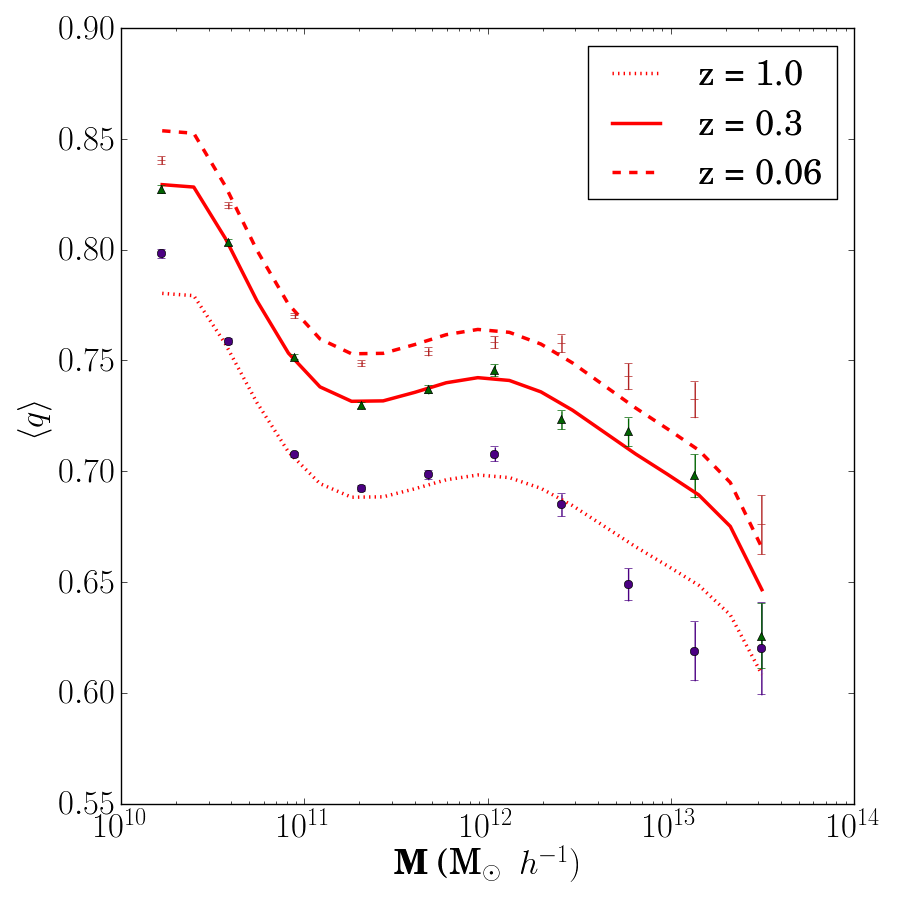} &
\includegraphics[width=3.2in,angle=0]{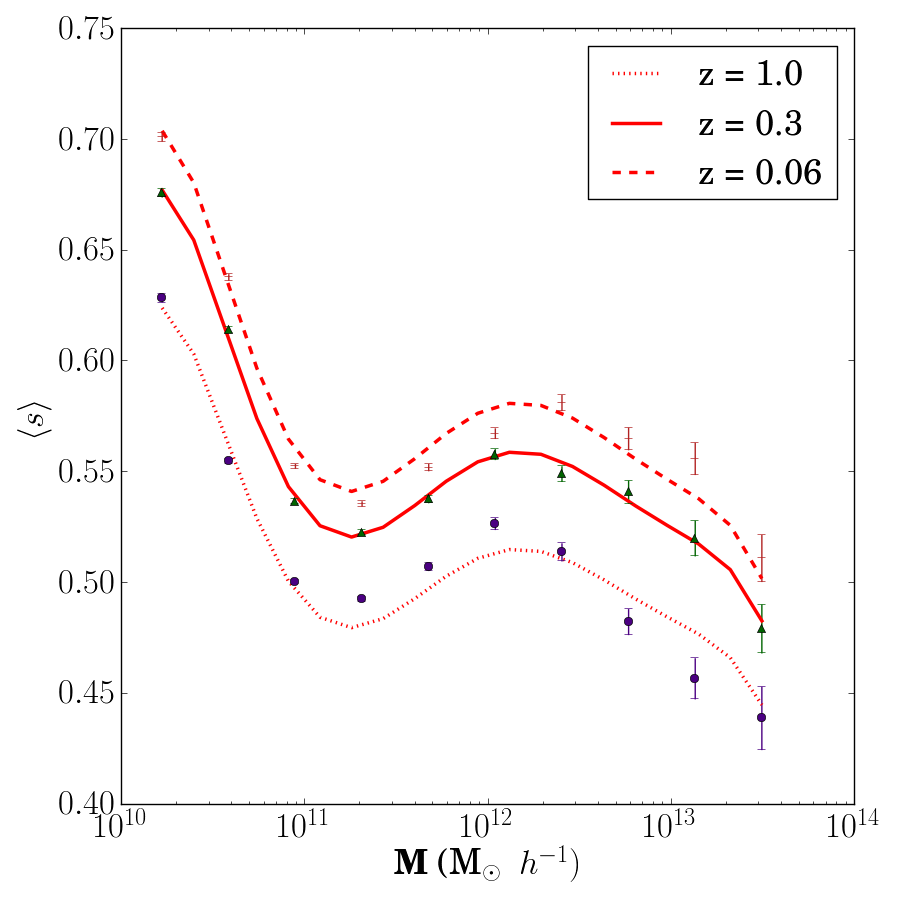}
\end{array}$
\caption{\label{F:fig_fit2}Fits for the axis ratios of shape defined by stellar matter in subhalos as a function of mass and redshift}
\end{center}
\end{figure*}


\begin{table*}
\begin{center}
\caption{\label{T:tab5} Parameters, $\gamma$ and $a_{i}$ for mean axis ratios, $ \langle q \rangle $ and $ \langle s \rangle $ in mass range, $10^{10.0} - 10^{14.0}$\hMsun}
\begin{tabular}{@{}lcccccccc}
\hline
Axis ratio  & $\gamma$ & $a_{0}$ & $a_{1}$ & $a_{2}$ & $a_{3}$ & $a_{4}$ & $a_{5}$ & $a_{6}$\\
\hline
$q$ (Dark Matter) & -0.12 & 0.797 & -0.049 & - & - & - & - & -\\
$s$ (Dark Matter) & -0.19 & 0.663 & -0.059 & - & - & - & - & -\\
$q$ (Stellar Matter) & -0.14 & 0.771 & -0.004 & -0.068 & -0.017 & -0.061 & -0.003 & -0.015 \\
$s$ (Stellar Matter) & -0.19 & -0.585 & 0.031 & -0.089 & -0.034 & 0.075 & -0.001 & -0.016 \\
\end{tabular}
\end{center}
\end{table*}


Here, we give the functional forms for the average axis ratios ($q$,
$s$) of shapes defined by dark matter and stellar matter in subhalos  as a function of mass and
redshift. The parameters are given in Table~\ref{T:tab5}. The plots showing fits for mean axis ratios of the shapes of dark matter and stellar
matter are given in Figs.~\ref{F:fig_fit1} and~\ref{F:fig_fit2}
respectively.

The fitting functions for average axis ratios are given by,
\begin{equation} \label{qsavgdmstr}
\langle q,s \rangle = (1+z)^{\gamma}\sum_{i}a_{i}\left[\log(\frac{M}{M_{piv}})\right]^{i}
\end{equation}
where, $M_{piv}$ is $10^{12}$\hMsun.

The fitting functions are linear in $\log(\frac{M}{M_{piv}})$ for shapes of dark matter with $i = {0,1}$ and polynomial to $6^{th}$ degree in $\log(\frac{M}{M_{piv}})$ with $i = {0,1,2,3,4,5,6}$ for shapes defined by stellar component in subhalos.
  
\section{Functional forms for probability distributions of 3D and 2D misalignment angles}
\label{appb}
\begin{figure}
\begin{center}
\includegraphics[width=3.2in,angle=0]{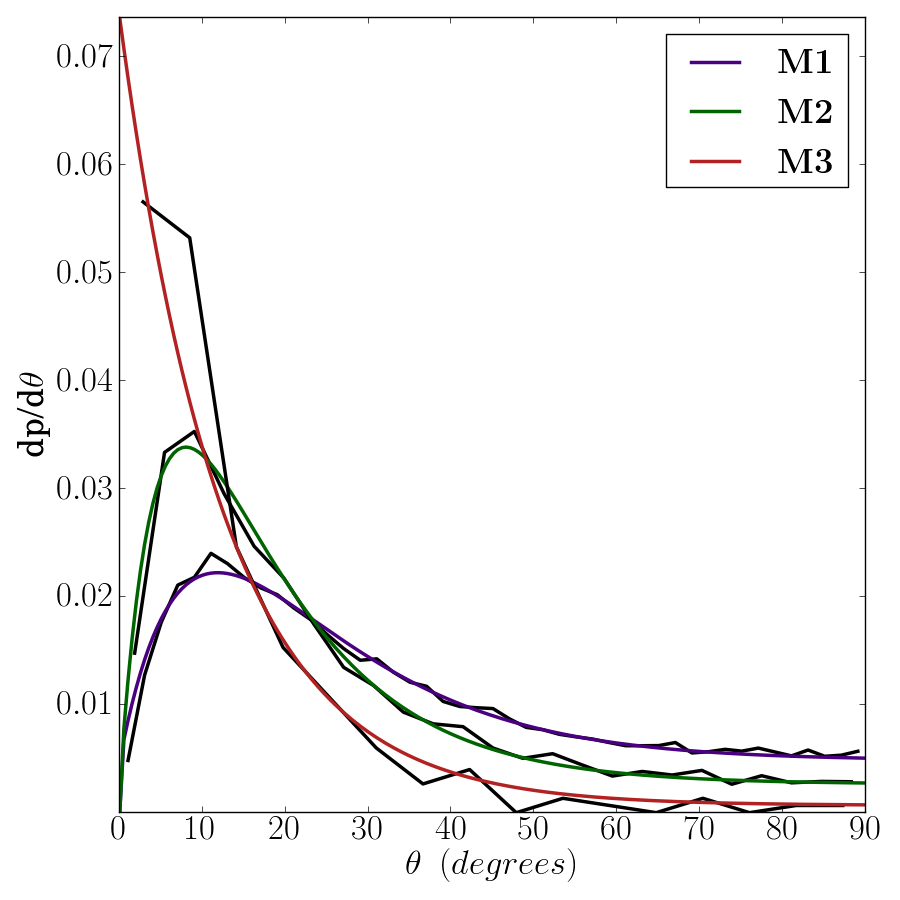} 
\caption{\label{F:fig_fit3a}Fits for probability distributions of 3D misalignment angles at $z = 0.3$}
\end{center}
\end{figure}

\begin{figure}
\begin{center}
\includegraphics[width = 3.2in]{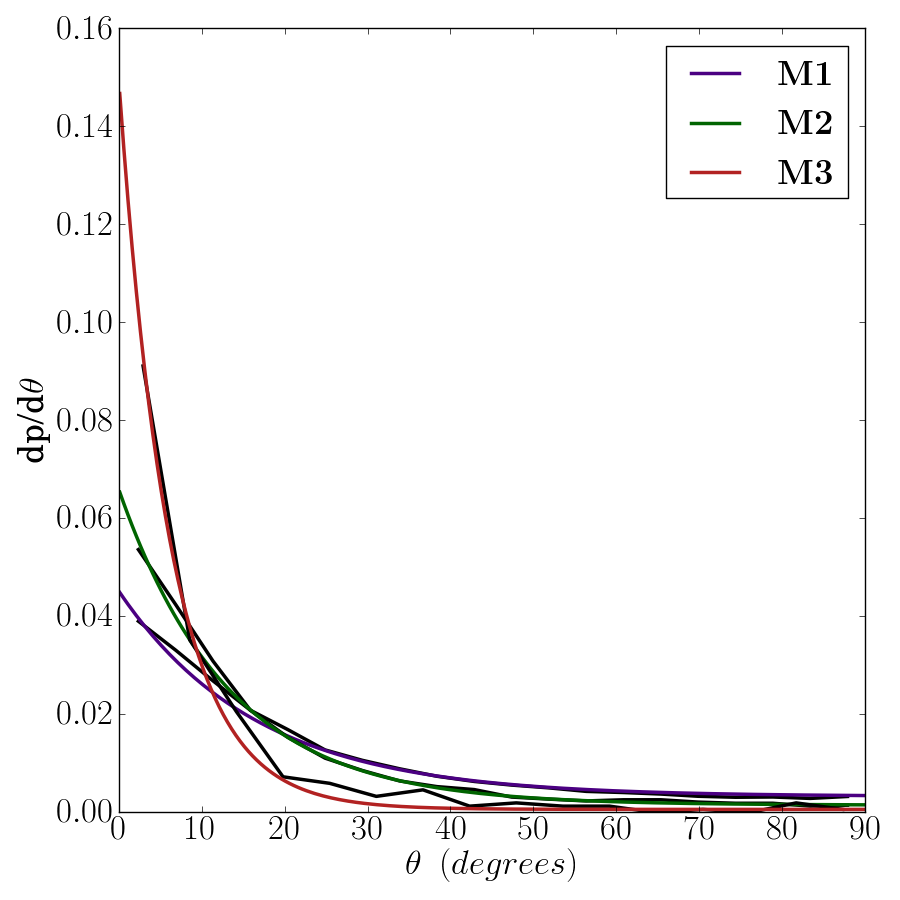}
\caption{\label{F:fig_fit3b}Fits for probability distributions of 2D misalignment angles at $z = 0.3$}
\end{center}
\end{figure}

\begin{table*}
\begin{center}
\caption{\label{T:tab7} Parameters for probability distributions of 3D misalignment angles at redshifts $z = 1.0, 0.3$, and $0.06$ for subhalos in the mass bins $M1: 10^{10.0}-10^{11.5}\hMsun, M2: 10^{11.5}-10^{13.0}\hMsun$ and $M3: ~> 10^{13.0}\hMsun$.}
\begin{tabular}{@{}lccccccccccccccc}
\hline
 & \multicolumn{5}{c}{$z = 1.0$} & \multicolumn{5}{c}{$z = 0.3$} & \multicolumn{5}{c}{$z = 0.06$} \\
\hline
Mass bin & $A_{z}$ & $B_{z}$ & $C_{z}$ & $\gamma_{z}$ & $\alpha_{z}$ & $A_{z}$ & $B_{z}$ & $C_{z}$ & $\gamma_{z}$ & $\alpha_{z}$ & $A_{z}$ & $B_{z}$ & $C_{z}$ & $\gamma_{z}$ & $\alpha_{z}$ \\
\hline
$ M1$ & $0.211$ & $0.079$ & $0.004$ & $0.023$ & $100$ & $0.146$ & $0.071$ & $0.005$ & $0.028$ & $100$ & $0.055$ & $0.052$ & $0.004$ & $0.071$ & $100$ \\
$ M2$ & $0.122$ & $0.088$ & $0.002$ & $0.134$ & $100$ & $0.091$ & $0.074$ & $0.003$ & $0.121$ & $100$ & $0.058$ & $0.057$ & $0.003$ & $0.166$ & $100$ \\
$ M3$ & $0.115$ & $0.119$ & $0.004$ & $-$ & $-$ & $0.073$ & $0.079$ & $-$ & $-$ & $-$ & $0.064$ & $0.070$ & $-$ & $-$ & $-$ \\
\end{tabular}
\end{center}
\end{table*}

\begin{table*}
\begin{center}
\caption{\label{T:tab8} Parameters for probability distributions of 2D misalignment angles at redshifts $z = 1.0, 0.3$, and $0.06$ for subhalos in the mass bins $M1: 10^{10.0}-10^{11.5}\hMsun, M2: 10^{11.5}-10^{13.0}\hMsun$ and $M3: ~> 10^{13.0}\hMsun$.}
\begin{tabular}{@{}lccccccccc}
\hline
 & \multicolumn{3}{c}{$z = 1.0$} & \multicolumn{3}{c}{$z = 0.3$} & \multicolumn{3}{c}{$z = 0.06$} \\
\hline
Mass bin & $A_{z}$ & $B_{z}$ & $C_{z}$ & $A_{z}$ & $B_{z}$ & $C_{z}$ & $A_{z}$ & $B_{z}$ & $C_{z}$ \\
\hline
$M1$ & $0.044$ & $0.060$ & $0.003$ & $0.042$ & $0.060$ & $0.003$ & $0.041$ & $0.056$ & $0.003$ \\
$M2$ & $0.077$ & $0.089$ & $0.001$ & $0.064$ & $0.075$ & $0.002$ & $0.056$ & $0.069$ & $0.002$ \\
$M3$ & $0.2$ & $0.211$ & $0.0$ & $0.146$ & $0.162$ & $0.0$ & $0.133$ & $0.137$ & $0.0$ \\
\end{tabular}
\end{center}
\end{table*}
The probability distributions for 3D misalignment angles in the two
lower mass bins $10^{10.0} - 10^{11.5}$\hMsun\ and $10^{11.5} -
10^{13.0}$\hMsun\ are given by
\begin{equation} \label{ma3d}
\frac{dp}{d\theta} = A_{z}(1 - e^{-\gamma_{z}\theta})e^{-B_{z}\theta} + (1 - e^{-\alpha_{z}\theta})C_{z}
\end{equation}
In the highest mass bin, $10^{13.0} - 10^{15.0}$\hMsun\ the fitting function is, 
\begin{equation} \label{ma3d3}
\frac{dp}{d\theta} = A_{z}e^{-B_{z}\theta}
\end{equation}

The probability distributions for 2D misalignment angles in different mass bins are given by
\begin{equation} \label{ma2d}
\frac{dp}{d\theta} = A_{z}e^{-B_{z}\theta} + C_{z}
\end{equation}

The fits for the probability distributions in 3D and 2D are shown in Fig.~\ref{F:fig_fit3a} and Fig.~\ref{F:fig_fit3b} respectively and the parameters are
given in Tables~\ref{T:tab7} and~\ref{T:tab8}.

\section{Functional forms for mean misalignment angles in 3D and 2D}
\label{appc}
\begin{figure}
\begin{center}
\includegraphics[width = 3.2in,angle=0]{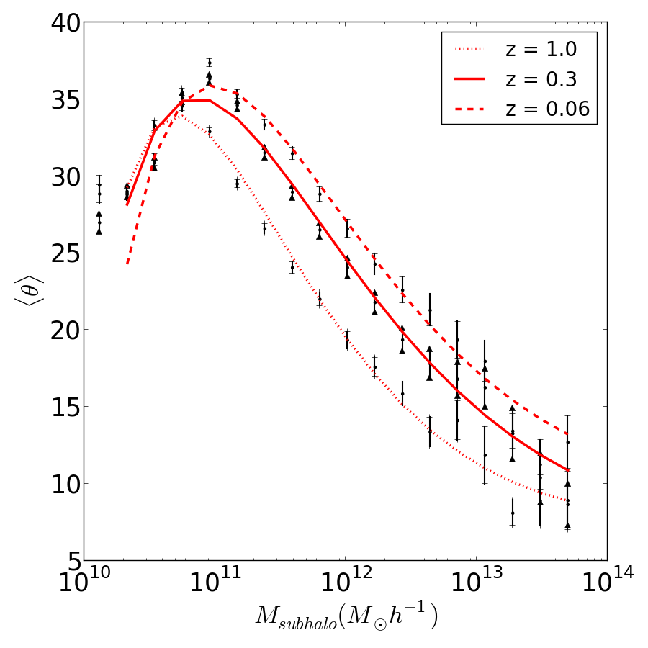} 
\caption{\label{F:fig_fit4a}Fits for mean misalignment angles in 3D as a function of mass}
\end{center}
\end{figure}
\begin{figure}
\begin{center}
\includegraphics[width = 3.2in,angle=0]{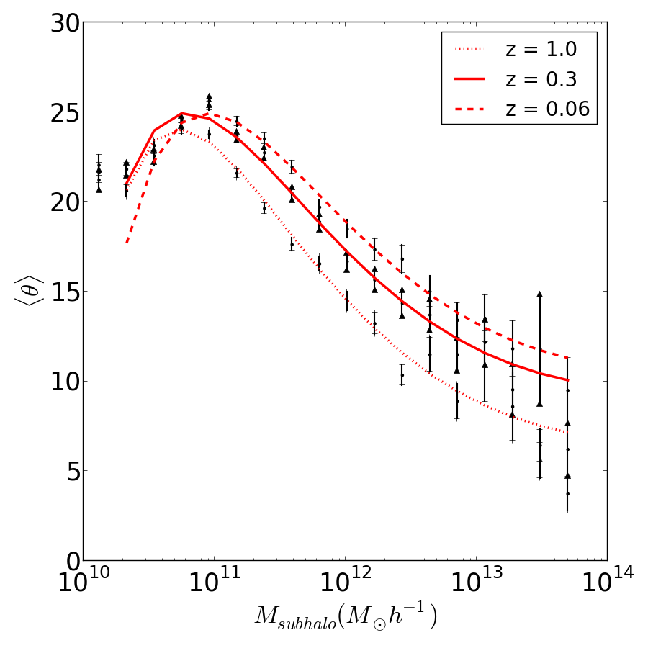}
\caption{\label{F:fig_fit4b}Fits for mean misalignment angles in 2D as a function of mass}
\end{center}
\end{figure}

\begin{table}
\caption{\label{T:tab9} Parameters for mean misalignment angles in 3D at redshifts $z = 1.0, 0.3$ and $0.06$ for subhalos in the mass range, $10^{10.0} - 10^{14.0}$ \hMsun.}
\begin{tabular}{@{}lcccccc}
\hline
$z$ & $a_{0z}$ & $a_{1z}$ & $b_{0z}$ & $c_{0z}$ & $c_{1z}$ & $d_{0z}$ \\
\hline
$1.0$ & $1.19$ & $-64.35$ & $0.79$ & $1.18$ & $-11.70$ & $10.19$ \\
$0.3$ & $0.88$ & $-53.72$ & $0.97$ & $1.16$ & $-11.46$ & $10.27$\\
$0.06$ & $1.09$ & $-28.58$ & $0.96$ & $1.38$ & $-13.74$ & $10.84$\\
\end{tabular}
\end{table}
\begin{table}
\caption{\label{T:tab10} Parameters for mean misalignment angles in 2D at redshifts $z = 1.0, 0.3$ and $0.06$ for subhalos in the mass range, $10^{10.0} - 10^{14.0}$ \hMsun.}
\begin{tabular}{@{}lcccccc}
\hline
$z$ & $a_{0z}$ & $a_{1z}$ & $b_{0z}$ & $c_{0z}$ & $c_{1z}$ & $d_{0z}$ \\
\hline
$1.0$ & $1.44$ & $-89.84$ & $0.82$ & $0.79$ & $-7.77$ & $9.92$ \\
$0.3$ & $1.86$ & $-79.50$ & $0.89$ & $0.89$ & $-8.75$ & $9.84$\\
$0.06$ & $2.07$ & $-43.40$ & $0.91$ & $0.90$ & $-8.99$ & $10.47$\\
\end{tabular}
\end{table}
The mean misalignment angles in 3D and 2D are given by,
\begin{equation} \label{avg3dma}
\theta(M) = (a_{0z} - a_{1z}e^{-(\frac{\log(M)-d_{0z}}{b_{0z}})})(c_{0z}\log(M) + c_{1z})
\end{equation}

The plots showing fits for mean misalignments in 3D and 2D are shown in Fig.~\ref{F:fig_fit4a} and Fig.~\ref{F:fig_fit4b} respectively. The corresponding parameters are given in Tables~\ref{T:tab9} and~\ref{T:tab10}.
\end{document}